\documentclass[aps,prb,twocolumn,showpacs,preprintnumbers,amsmath,floatfix]{revtex4-1}
\usepackage{graphicx}
\usepackage{dcolumn}
\usepackage{subfigure}
\usepackage{wrapfig}
\usepackage{cancel}
\usepackage{color}
\usepackage{bm}

\begin{document}

\title{\bf  Spin fluctuations and superconductivity in a 3D
tight-binding model for BaFe$_2$As$_2$}
\author{S. Graser$^1$, A. F. Kemper$^2$, T. A. Maier$^3$, H.-P. Cheng$^2$,
P. J. Hirschfeld$^2$, and D. J. Scalapino$^4$}
\affiliation{$^1$Center for Electronic Correlations and Magnetism,
Institute of Physics, University of Augsburg, D-86135 Augsburg, Germany\\
$^2$Department of Physics, University of Florida,
Gainesville, FL 32611, USA\\$^3$ Center for Nanophase Materials
Sciences and Computer Science and Mathematics Division, Oak Ridge
National Laboratory, Oak Ridge, TN 37831-6494
\\$^4$ Department of Physics, University of California, Santa
Barbara, CA 93106-9530 USA}
\date{\today}

\begin{abstract}
Despite the wealth of experimental data on the Fe-pnictide
compounds of the $K$Fe$_2$As$_2$-type, $K$ = Ba, Ca, or Sr,
the main theoretical work
based on multiorbital tight-binding models has been restricted so far to the study
of the related 1111 compounds.
This can be ascribed to the more three dimensional electronic structure found by 
{\it ab initio} calculations for the 122
materials,  making this system less amenable to
model development. In addition, the
more complicated Brillouin zone (BZ) of the body-centered tetragonal
symmetry does not allow  a straightforward unfolding of the electronic
band structure into an effective 1Fe/unit cell BZ.
Here we present an effective 5-orbital tight-binding fit of the full
DFT band structure for BaFe$_2$As$_2$ including the $k_z$ dispersions.
We compare the 5-orbital spin fluctuation model to  one previously studied for LaOFeAs
and calculate the RPA enhanced susceptibility. Using the fluctuation exchange 
approximation to determine the leading pairing instability, we then examine
the differences between a strictly two dimensional model calculation   over a single
$k_z$ cut of the BZ  and a completely three dimensional approach.
We find pairing states quite similar to the 1111 materials, with generic quasi-isotropic pairing
on the hole sheets and nodal states on the
electron sheets at $k_z=0$ which however are gapped as the system is hole doped.
On the other hand, a substantial
$k_z$ dependence of the order parameter remains, with most of the pairing strength 
deriving from processes near $k_z=\pi$. These states exhibit a tendency for an
enhanced anisotropy on the hole sheets and a reduced
anisotropy on the electron sheets near the top of the BZ.
\end{abstract}


\maketitle

\section{ Introduction}
\label{sec:intro}

The discovery of superconductivity in the Fe-based pnictide and
chalcogenide compounds has stimulated a tremendous research effort
in many areas of solid state physics and chemistry.  
The materials  initially discovered by the Hosono group, 
LaOFeP\cite{ref:Kamihara2006} and fluorine doped LaOFeAs~\cite{ref:Kamihara2008}, 
belong to a class of iron pnictides that are commonly referred
to as the 1111 structures. More recently, a great deal of attention has  been
devoted to the preparation and investigation of materials where the
FeAs layers are separated by a single cation only, the so-called
122 structures~\cite{ref:Rotter}. Here both hole-doping by replacing
Ba in part by K as well as electron-doping due to a fractional substitution
of Fe by Co have proven successful in suppressing the spin-density wave (SDW)
formation in favor of a superconducting ground state~\cite{ref:Rotter,ref:Sefat}.
Although the maximum critical temperature of $T_c=38 K$ in the 122-systems
is smaller than in the related 1111-materials, the
possibility of growing high quality single crystals with relatively 
good surfaces make them optimal candidates for comprehensive
experimental studies. Angle-resolved photoemission electron spectroscopy
(ARPES) measurements performed on high quality single crystals of
Ba$_{0.6}$K$_{0.4}$Fe$_2$As$_2$~\cite{ref:Zhao,ref:Ding,ref:Kondo,ref:Evtushinsky,ref:Nakayama,ref:Hasan} 
have been very influential, revealing the position,
shape and size of the Fermi surface pockets that are in qualitative agreement
with band structure calculations~\cite{ref:Singh,ref:Sefat}.
In addition, the ARPES experiments also claim to resolve
the size and momentum space distribution
of the superconducting gap, showing at least two distinct values
of the order parameter and a nearly isotropic gap size distribution
along the individual Fermi surface sheets.
These observations can in principle be understood in terms of the
formation of a sign-changing $s$-wave state generated by the
exchange of spin fluctuations~\cite{ref:Mazin,ref:Dong}.
Such a gap structure is also supported by neutron scattering 
experiments on these compounds, which find a resonance emerging 
in the superconducting state at a wave vector that corresponds to 
the separation between hole and electron pockets~\cite{ref:Christianson,
ref:Lumsden,ref:Chi,ref:Li,ref:Inosov}.
Despite these promising results, the symmetry of the superconducting 
order parameter in these materials is still controversial, 
and many experiments imply the existence of low-energy quasiparticle 
excitations, denoting the possible existence of nodes.
These include NMR~\cite{ref:RKlingeler,ref:Zheng,ref:Grafe,ref:Ahilan,ref:Nakai,ref:MYashima2009}, 
superfluid density~\cite{ref:Hashimoto,ref:Malone_Martin,ref:Hashimoto2,ref:Gordon,ref:Gordon2,ref:Fletcher}, 
thermal conductivity\cite{ref:Tailleferkappa,ref:TanatarCo2009}
and Raman light scattering~\cite{ref:Muschler}.  
The existence of nodes is also suggested by multiorbital calculations for
the 1111 materials~\cite{ref:Kuroki,ref:GraserNJP}, although a transition 
from such a nodal state to a fully gapped state has also been described in such
a framework~\cite{ref:Kuroki2,ref:chubukov_nodal_gapped,ref:bernevig_nodal_gapped,
ref:Thomale_nodal_gapped,ref:Wang_nodal_gapped,ref:kemper_nodal_gapped}.
While these works indicate a sensitivity of the superconducting state to details of the
electronic structure,  particularly the position of the pnictogen, they have not been 
discussed in the framework of the 122 materials.  

In particular, it is important to ask if the greater three-dimensionality of the
electronic structure in these materials has qualitative effects on the pair state.
Certainly the effects of near nesting of the hole and electron sheets of
 the Fermi surface, which are said to
be responsible for the stabilization of the $s_\pm$ state, must be expected to change for those
sheets which are strongly $k_z$-dispersive.
One important aspect in this regard is the role of doping since 
different impurities may not only provide carriers for the FeAs planes
but can also alter the hybridization of states between layers, as argued for
the case of Co doping by Kemper {\it et al.}~\cite{ref:Kemper}.
In general, the more three dimensional character and the
resulting coupling between the individual FeAs
layers in the 122 compounds requires us to revisit
the existing models and to reassess the universality
of the results obtained so far.

\section{Tight-binding fit of the LDA band structure}
\label{sec:TB}

\begin{figure}
\begin{flushleft}
(a) \hspace{0.45\columnwidth} (b)
\end{flushleft}
\vspace{-0.5cm}
\includegraphics[width=.49\columnwidth]{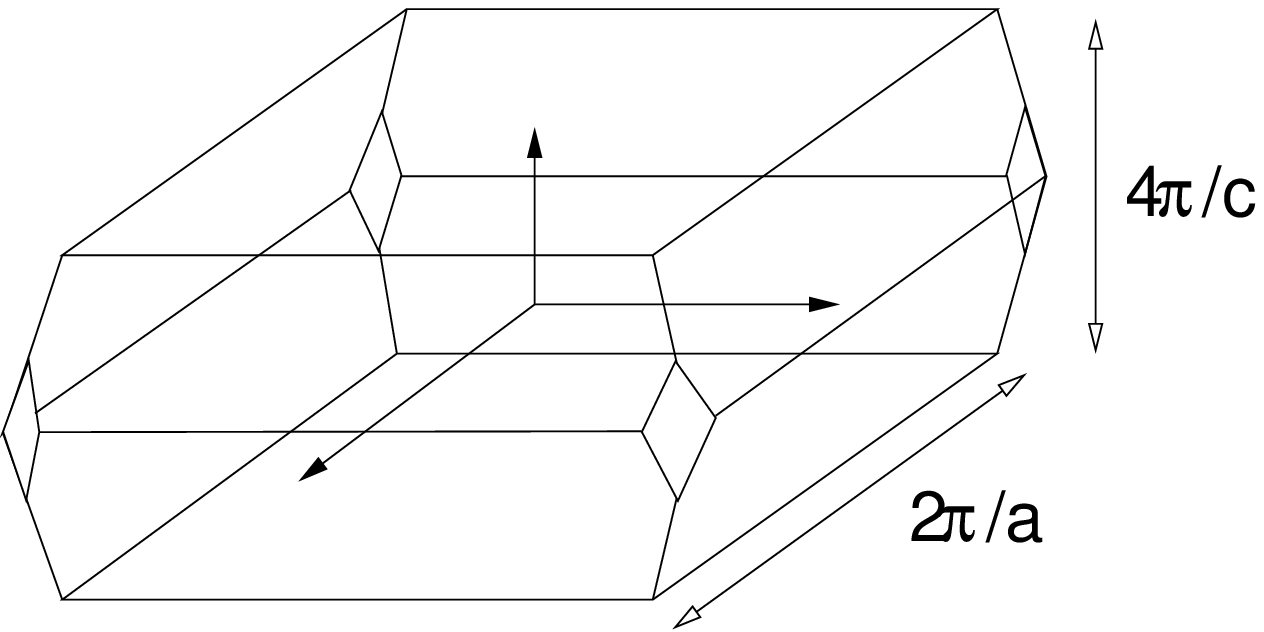}
\includegraphics[width=.49\columnwidth]{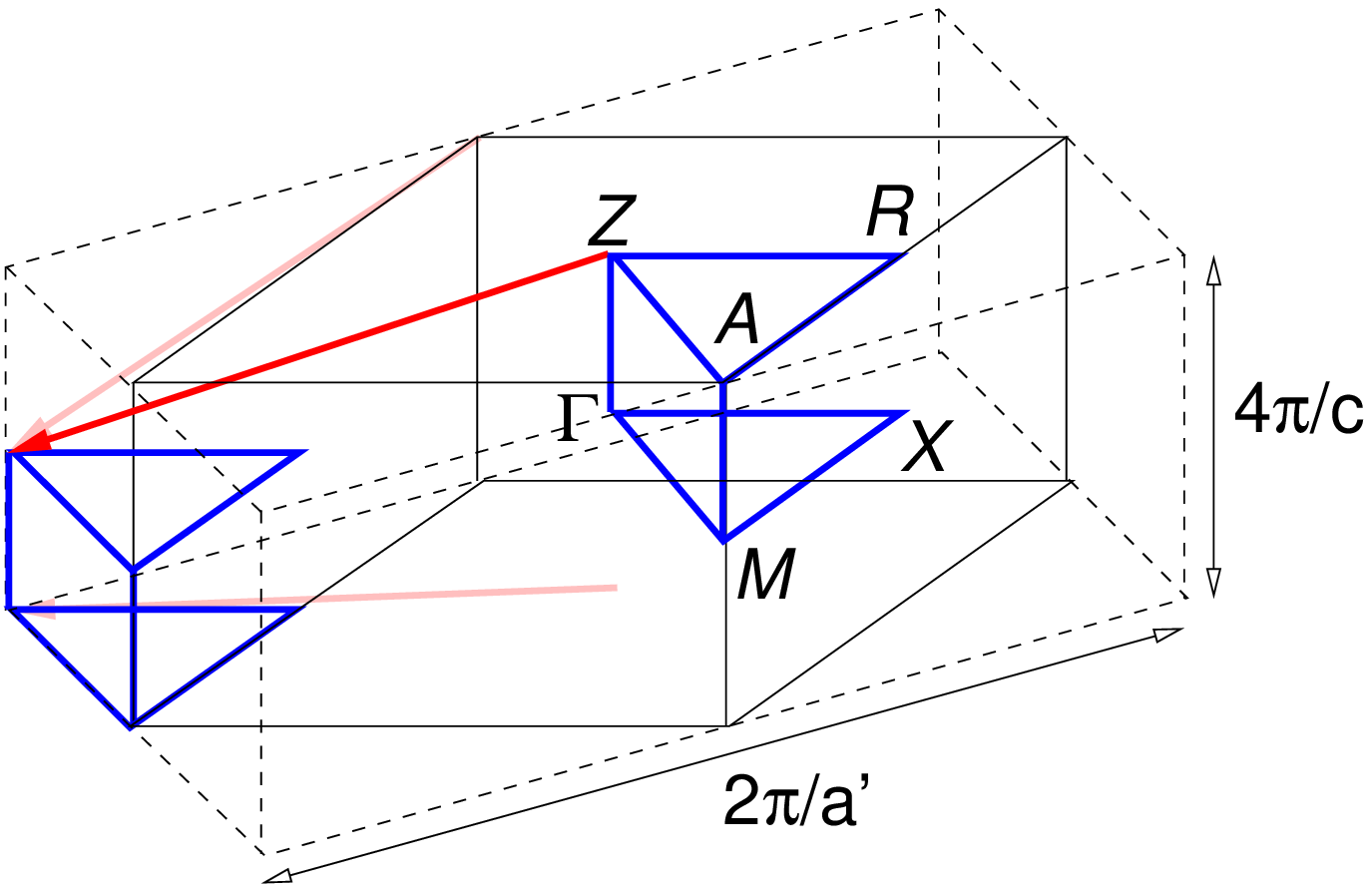}
\caption{(Color online) Sketch of the Brillouin zone of the
{\it I}4/{\it mmm} crystal symmetry (a) and of the large effective
BZ corresponding to the 1Fe/unit cell (b).
The blue line shows the two paths in the 1Fe/unit cell
BZ that have to be folded by the reciprocal lattice
vector ${\mathbf T} = (\pi,\pi,\pi)$ (red arrow) to give
the corresponding path in the 2Fe/unit cell BZ of the
{\it P}4/{\it nmm} symmetry. 
}
\label{fig:BZ}
\end{figure}

Compared to the other pnictide and chalcogenide
superconductors as e.g. the 1111 materials (LaOFeAs),
the 111 systems (LiFeAs), or the binary 11 compounds (FeSe),
there is no straightforward way to describe the
electronic structure of the 122 materials (BaFe$_2$As$_2$)
in an effective 5-orbital Fe model.
This has several reasons: first of all the Ba atoms, forming
a spacing layer between the FeAs planes, contribute significantly
to the interlayer hopping. Secondly, we have only one Ba atom per
unit cell, but two Fe and two As atoms, therefore a description in
an effective model based on a 1Fe/unit cell BZ can only be successful
if the Ba bands are integrated out from the beginning. Finally the
basic unit cell in the 122 materials is not a simple tetragonal unit cell
but a body-centered tetragonal unit cell, with non-Cartesian reciprocal
lattice vectors and a complicated Brillouin zone (see Fig.~\ref{fig:BZ}).
Despite these difficulties it is in principle possible to define
an effective band structure in the 1Fe/unit cell BZ starting from a purely
Fe-based fit of the full DFT band structure~\cite{ref:Miyake}.

We have calculated the band structure for the BaFe$_2$As$_2$
parent compound making use of the density functional theory (DFT)
in a plane wave basis set with ultrasoft pseudopotentials
as provided in the Quantum ESPRESSO package~\cite{ref:QE}.
Here we have used the lattice constants as well as the internal coordinates
tabulated in Ref.~\onlinecite{ref:RotterSDW} with $a=3.9625$~\AA,
$c = 13.0168$~\AA, and $z_{As} = 0.3545$.
\begin{figure}
\begin{flushleft}
(a)  \\
\end{flushleft}
\vspace{-0.5cm}
\includegraphics[width=0.9 \columnwidth]{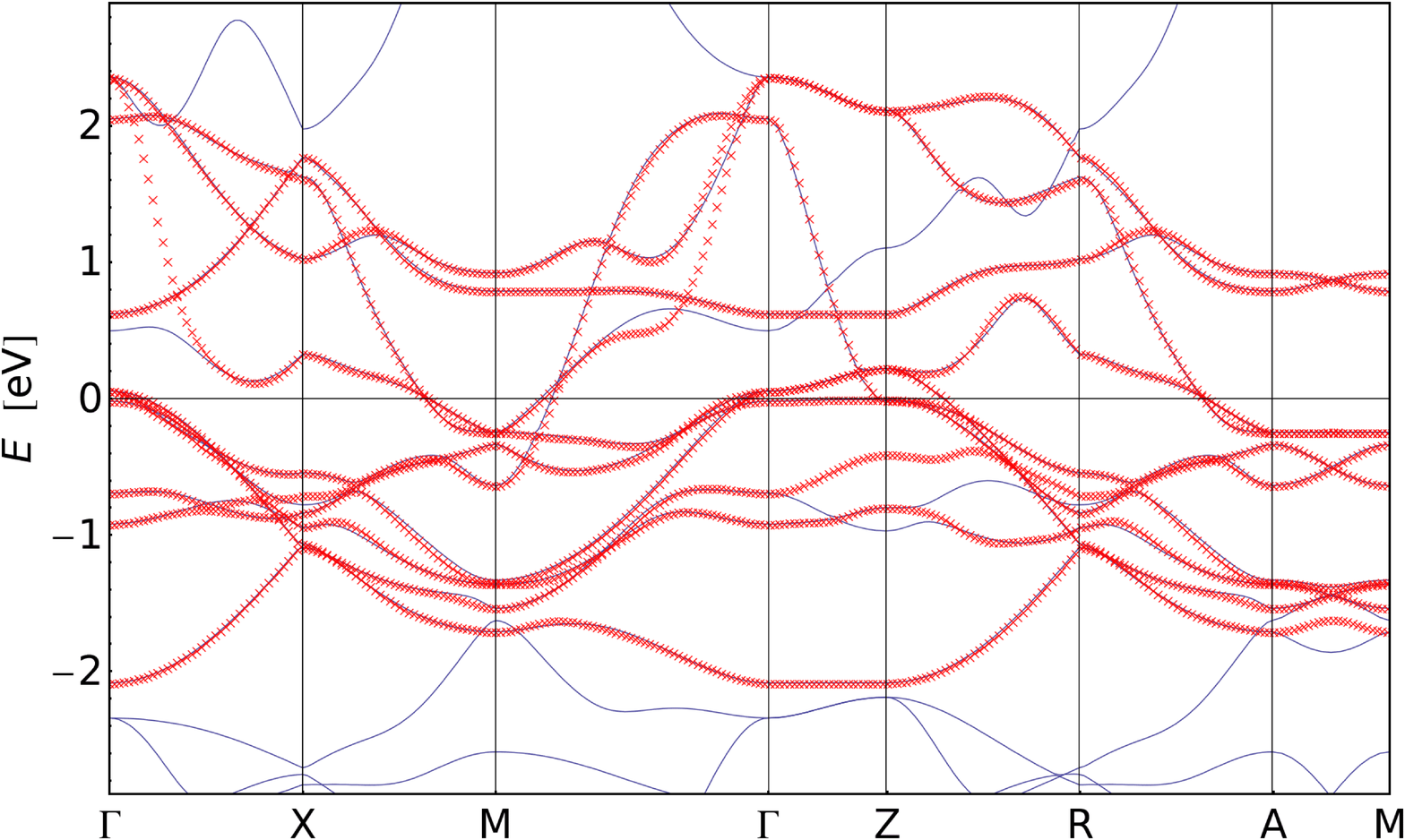} \\
\begin{flushleft}
(b)  \\
\end{flushleft}
\vspace{-0.5cm}
\includegraphics[width=0.9 \columnwidth]{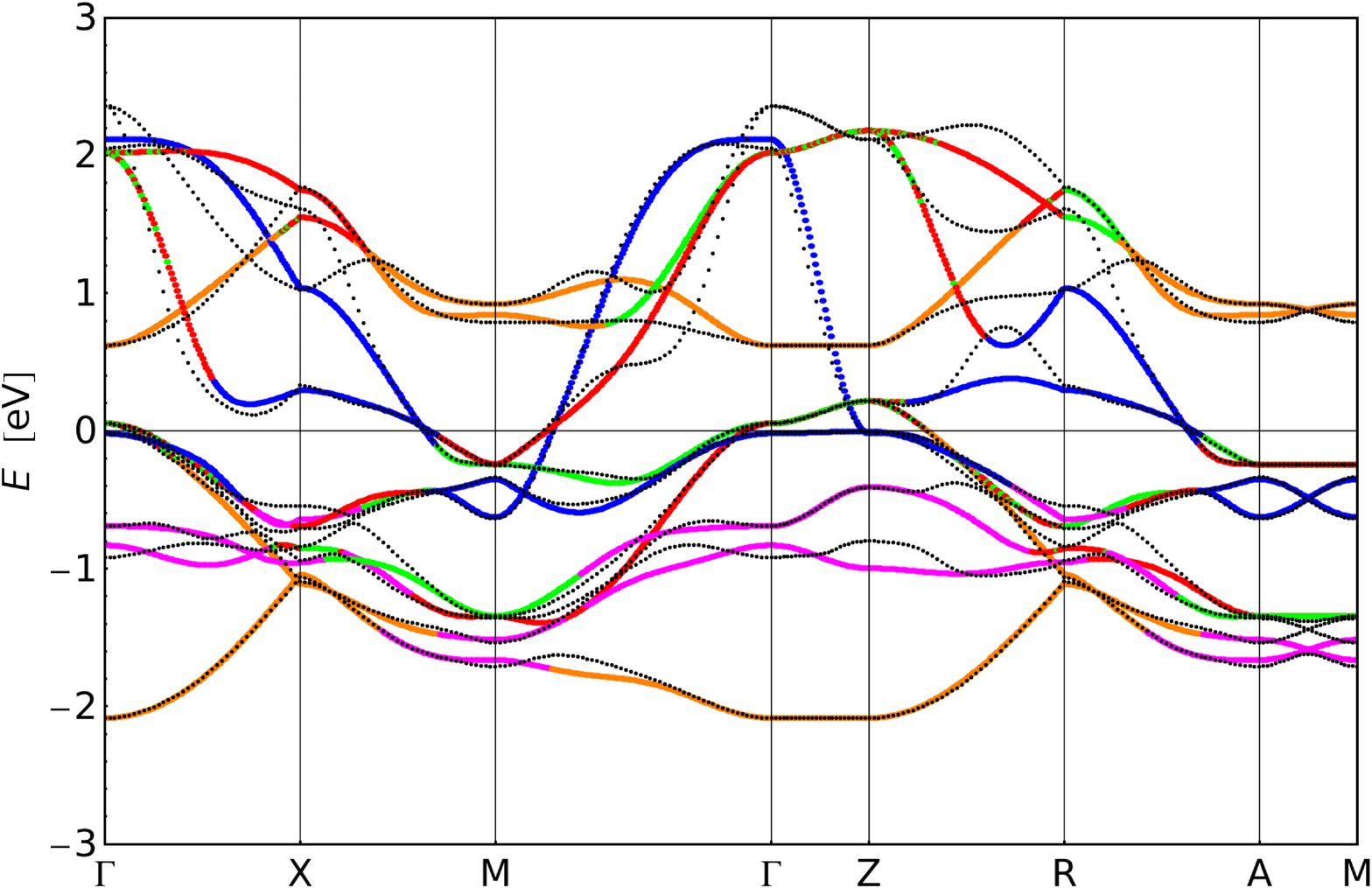}
\caption{(Color online) The paramagnetic DFT band structure (full line) and a Wannier fit (crosses) of the
10 bands in the vicinity of the Fermi surface onto the Fe-$3d$ orbitals (a).
The 5-orbital tight-binding fit (colored points)
of the 10-orbital Wannier fit (black points) with a color coding of the main orbital
contributions (b). The colors correspond to $d_{xz}$ (red), $d_{yz}$ (green),
$d_{xy}$ (blue), $d_{x^2-y^2}$ (orange), and $d_{3z^2-r^2}$ (magenta).}
\label{fig:bands}
\end{figure}
\begin{figure}
\includegraphics[width=0.8 \columnwidth]{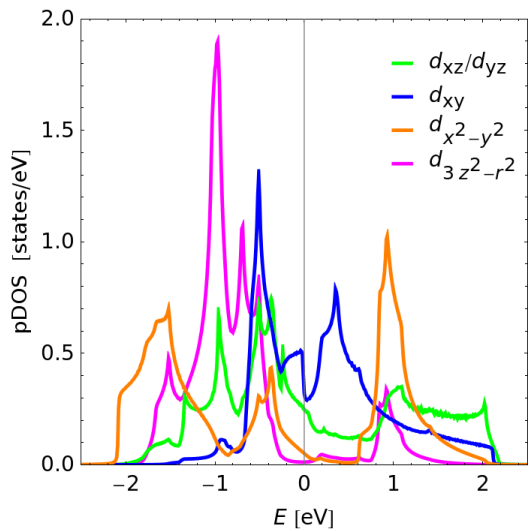} 
\caption{(Color online) The partial density of states of the 
5-orbital tight-binding fit, using the same color coding as in Fig.~\ref{fig:bands}
b.}
\label{fig:pDOS}
\end{figure}

\begin{figure}
\begin{flushleft}
\hspace{0.05\columnwidth} (a)
\end{flushleft}
\includegraphics[width=0.9 \columnwidth]{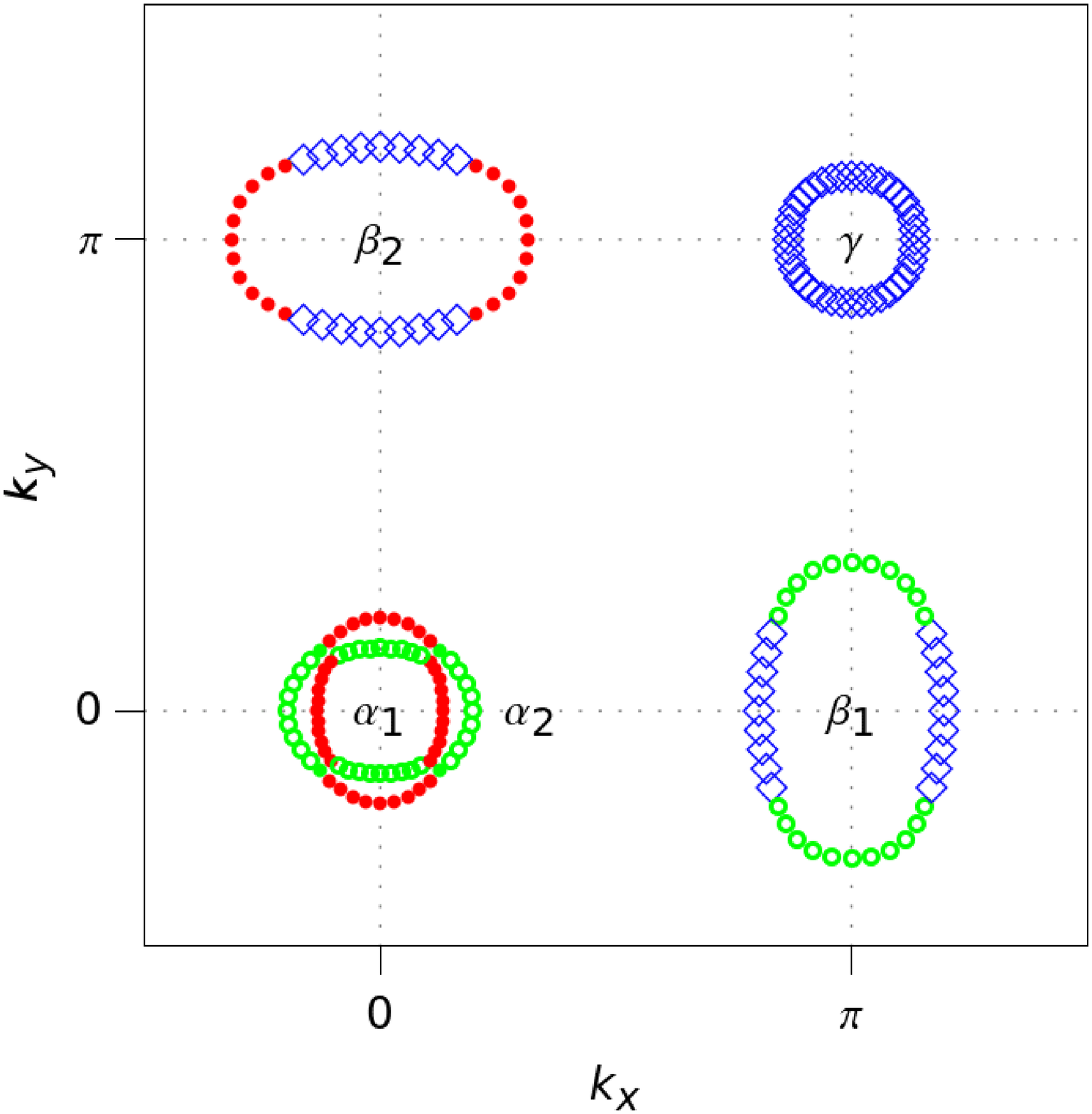}
\begin{flushleft}
\hspace{0.05\columnwidth} (b)
\end{flushleft} 
\includegraphics[width=0.9 \columnwidth]{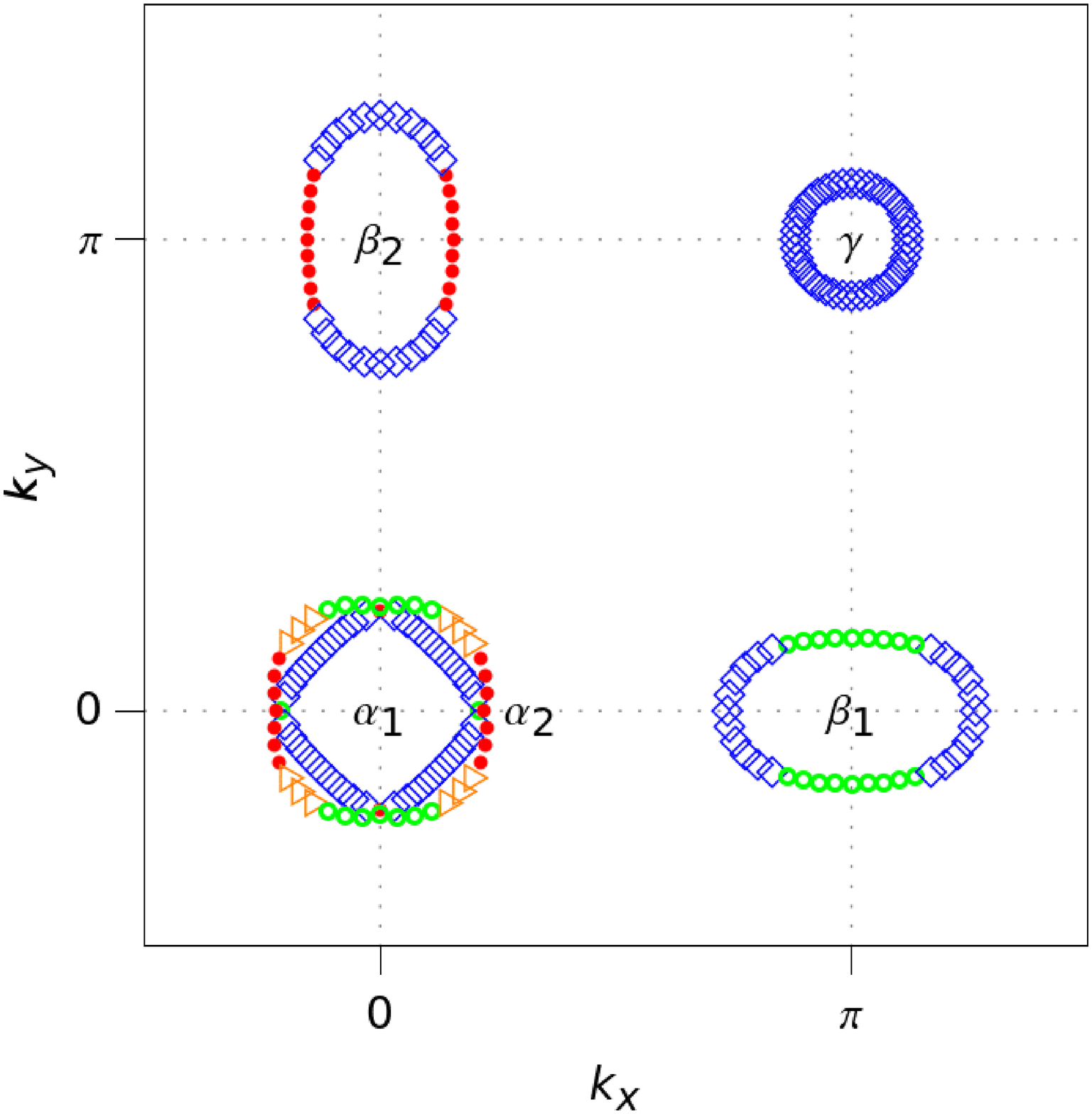}
\caption{(Color online) The main orbital contributions to the Fermi surfaces 
of the hole doped compound with $\langle n \rangle = 5.9$
at $k_z=0$ (a) and $k_z=\pi$ (b) using the same color coding as in 
Fig.~\ref{fig:bands} b.}
\label{fig:FS}
\end{figure}

The calculations were performed for the body-centered tetragonal unit cell
corresponding to the {\it I}4/{\it mmm} symmetry of the crystal
but we have plotted the bands along the high symmetry lines
of a corresponding simple tetragonal unit cell to facilitate
the comparison with the band structure of the 1111 materials.
In the next step, we projected the bands in the vicinity
of the Fermi energy on the Fe-$3d$ orbitals using maximally
localized Wannier functions (MLWF) following the method of
Marzari and Vanderbilt~\cite{ref:Marzari}.
The bands were disentangled by minimizing the spread
of the Wannier functions. Except for a band with mainly Ba
character that approaches the Fermi energy between
$\Gamma$ and $X$ this projection reproduces the full
DFT band structure very accurately as can be seen from Fig.~\ref{fig:bands} a.
Finally, we fitted the Wannier bands with a 5-orbital tight-binding Hamiltonian,
unfolding the small 2Fe/unit cell BZ to a large 1Fe/unit cell effective BZ.
The Hamiltonian is given as
\begin{equation}
H_0 = \sum_{{\bf k}\sigma} \sum_{mn} \left( \xi_{mn} ({\bf k}) + \epsilon_m \delta_{mn} \right)
d_{m\sigma}^\dagger({\bf k}) d_{n\sigma}({\bf k})
\label{eq:Ham}
\end{equation}
where $d_{m,\sigma}^\dagger({\bf k})$ creates a particle with momentum
${\bf k}$ and spin $\sigma$ in the orbital $m$. The kinetic energy terms
$\xi_{mn} ({\bf k})$ together with the parameters of the 5-orbital
tight-binding fit are listed in the appendix. The orbital resolved 
density of states shown in Fig.~\ref{fig:pDOS} reveals dominant 
contributions to the total density of states at the Fermi level 
from the Fe $d_{xz}/d_{yz}$ and $d_{xy}$ orbitals.
\begin{figure}
\begin{flushleft}
\hspace{0.05\columnwidth} (a)
\end{flushleft}
\includegraphics[width=0.9 \columnwidth]{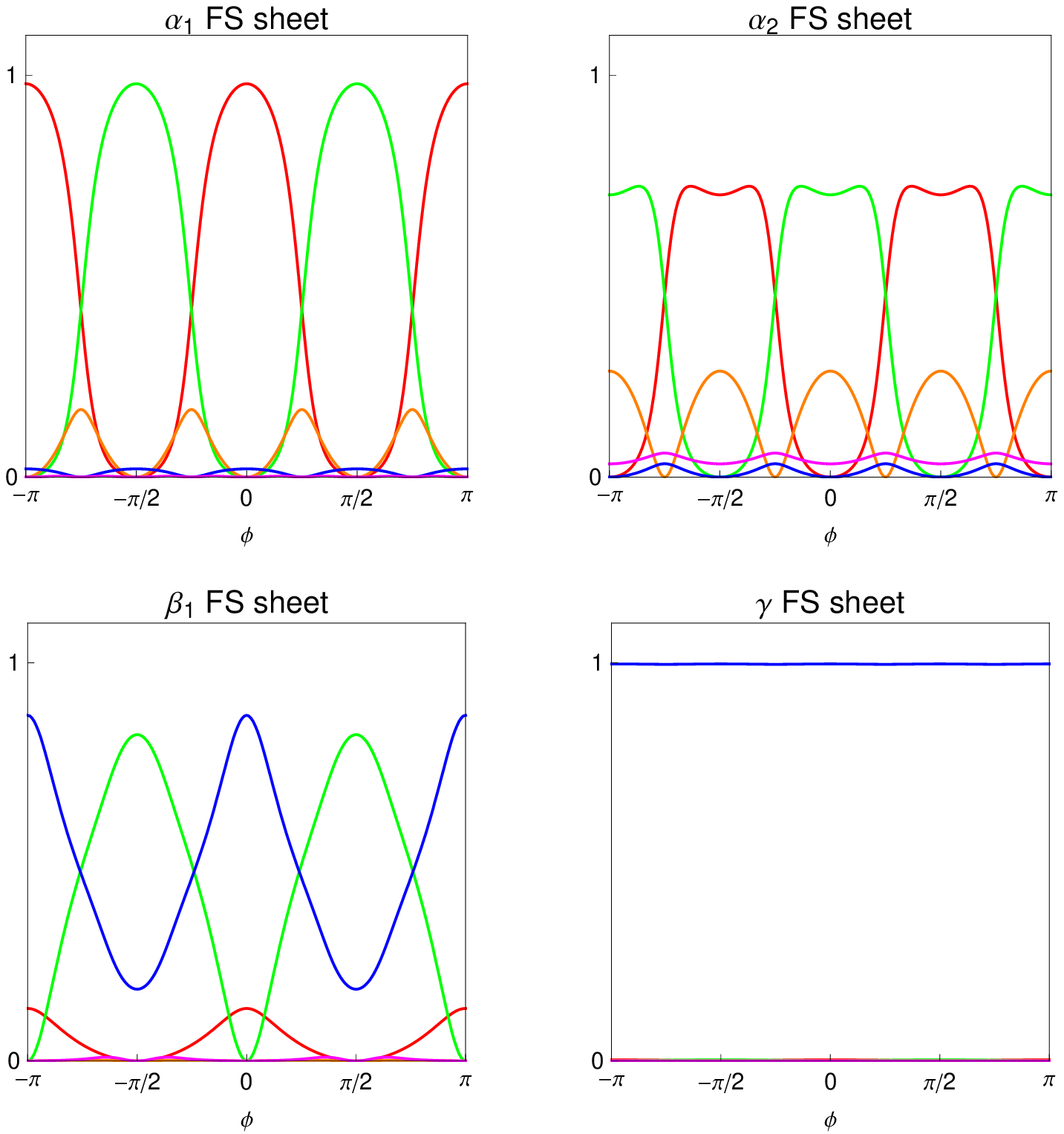}
\begin{flushleft}
\hspace{0.05\columnwidth} (b)
\end{flushleft} 
\includegraphics[width=0.9 \columnwidth]{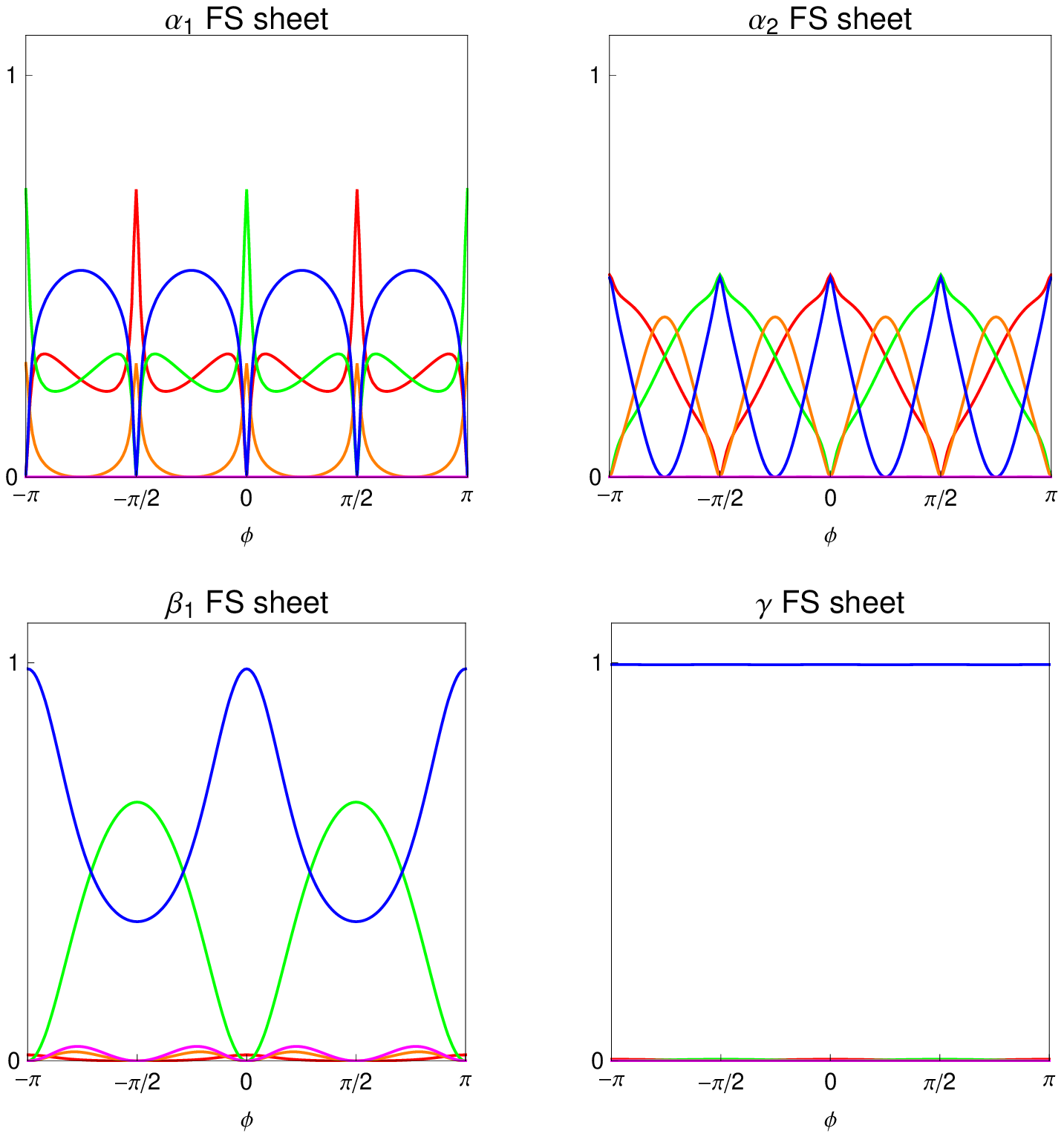}
\caption{(Color online) The orbital composition of the Fermi surface sheets
for a hole doped compound ($\langle n \rangle=5.9$) as given by $|a_{\nu}^t({\bf k})|^2$ for $k_z=0$ (a) 
and $k_z=\pi$ (b). The orbital
contributions are shown as a function of the winding angle $\alpha$ 
starting with the rightmost point on each Fermi surface
sheet.}
\label{fig:OC}
\end{figure} 
In Fig.~\ref{fig:FS} we show the Fermi surface pockets 
at two different $k_z$ cuts of the BZ,
where the colors encode the main orbital contributions to the
respective band. As discussed in Kemper {\it et al.}~\cite{ref:kemper_nodal_gapped}, 
the important orbital matrix elements in the pairing interaction 
enter as $|a^t_{\nu}({\bf k})|^2$, where $t$ denotes the orbital and $\nu$ the band index.  
Plots of $|a^t_{\nu}({\bf k})|^2$ for $k_z=0$ and $k_z=\pi$ are shown in Fig.~\ref{fig:OC} a and b,
respectively. This figure is 
similar to Fig.~\ref{fig:FS}, but contains more detailed information, since it shows all 
of the orbital contributions, while Fig.~\ref{fig:FS} shows only the largest orbital 
contribution on a given part of each Fermi surface. Here we note that 
the orbital composition of the Fermi surfaces at $k_z=0$ is similar to the one found for the 1111 materials
(e.g. compare with Fig. 5b in Ref.~\onlinecite{ref:GraserNJP}), while
at $k_z=\pi$ the orbital composition changes substantially.
Here the inner hole pocket around $\Gamma$ (labeled as $\alpha_1$)
is of predominantly $d_{xy}$ character while the outer
hole pocket around $\Gamma$ (labeled as $\alpha_2$) has in addition to the 
$d_{xz/yz}$ contributions significant involvement of the 
$d_{x^2-y^2}$ orbitals. From $k_z=0$ to $k_z=\pi$
one also finds a change in the orientation
of the ellipticity of the electron pockets $\beta_1$ and $\beta_2$
without a qualitative change of the orbital composition. 
Recently, a similar procedure was used by Miyake {\it et al.}~\cite{ref:Miyake} 
to determine an effective tight-binding model $\xi_{mn}({\bf k})$ for BaFe$_2$As$_2$ 
based on a DFT calculation. 
Their results appear to agree roughly with ours, with the
exception that they used a different choice for the
phase of the orbital basis, leading to a discrepancy
in the sign of some of the dispersions.  This difference
should not affect the eigenenergies, however.

\begin{figure}
\includegraphics[width=.9\columnwidth]{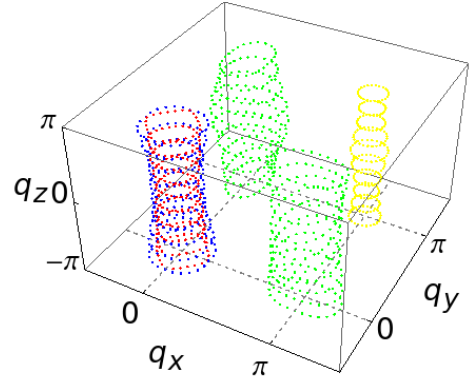}
\caption{(Color online) Fermi surface mesh of the hole doped 
compound applied to the calculation of the pairing functions.
Here we used $24\times 10$ ${\bf k}$-points for every Fermi surface sheet
with $\alpha_1$ (red), $\alpha_2$ (blue), $\beta_1$, $\beta_2$ (green), and
$\gamma$ (yellow).}
\label{fig:FermiS}
\end{figure}

\section{3D multiorbital susceptibility}
\label{sec:susc}

In this section we discuss the differences between a
susceptibility calculated for a fixed value of $k_z$ neglecting
the $k_z$ dispersion of the energy bands and a complete 3D calculation of the
susceptibility, taking the full momentum dependence of the
band structure into account. 
Using the notation of Ref.~\onlinecite{ref:Kubo} we write
the non-interacting susceptibilities as
\begin{equation}
\chi_{qtsp} ({\bf q},i\omega_m) = - \frac{1}{N\beta} \sum_{{\bf k},i\omega_n} G_{pt} ({\bf k},i\omega_n)
G_{qs} ({\bf k}+{\bf q},i\omega_n+i\omega_m)
\end{equation}
where $N$ is the number of Fe lattice sites, $\beta=1/T$ is the inverse temperature,
$\omega_n$ are the fermionic and $\omega_m$ the bosonic Matsubara
frequencies in the imaginary time formalism, and $s$, $t$, $p$, and $q$
are indices denoting the Fe-$3d$ orbitals. For the full 3D susceptibility
the momentum sum runs over $k_x$, $k_y$, and $k_z$, while for the 2D calculations
$k_z$ is kept fixed and the susceptibility is only evaluated at $q_z=0$.
For the integration we use a $64 \times 64 \times 20$ $k$-mesh and
we notice only negligible finite size effects, that show up as weak
oscillations at small ${\bf q}$.
The spectral representation of the Green's function is given as
\begin{equation}
G_{sp} ({\bf k},i\omega_n) = \sum_\mu \frac{a_\mu^s({\bf k}) 
a_\mu^{p*}({\bf k})}{i \omega_n - E_\mu({\bf k})}
\end{equation}
where the matrix elements $a_\mu^s({\bf k}) = \langle s | \mu {\bf k} \rangle$,
connecting the orbital and the band space, are determined by a diagonalization
of the  intra- and interorbital dispersions of the tight-binding
Hamiltonian given in Eq.~\ref{eq:Ham}. Now we calculate the retarded susceptibility as
\begin{eqnarray}
\chi_{qtsp} ({\bf q},\omega) & = & - \frac{1}{N} \sum_{{\bf k},\mu\nu}
\frac{a_\mu^p({\bf k}) a_\mu^{t*}({\bf k}) a_\nu^q({\bf k}+{\bf q}) a_\nu^{s*}({\bf k}+{\bf q})}
{ \omega + E_\nu({\bf k}+{\bf q}) - E_\mu({\bf k}) + i 0^+} \nonumber \\
& & \times \left[ f(E_\nu({\bf k}+{\bf q})) - f(E_\mu({\bf k})) \right]
\end{eqnarray}
It is evident that for a system without $k_z$ dispersion the susceptibility does not
depend on $q_z$ and the sum over $k_z$ can be neglected.
Finally, we take the orbital dependent interactions into account
by defining the RPA enhanced spin susceptibility as
\begin{equation}
(\chi_1^{\rm RPA})_{stpq} = \chi_{stpq} +
(\chi_1^{\rm RPA})_{stuv} (U^s)_{uvwz} \chi_{wzpq},
\end{equation}
where we sum over repeated indices. Here the interaction matrix
$U^s$ is nonzero only for
\begin{equation}
U_{aaaa}^s = \bar{U}, \, U_{bbaa}^s = \bar{J}, \,
U_{abab}^s = \bar{U}', \, U_{abba}^s = \bar{J}' \nonumber
\end{equation}
where $a\neq b$ and the definitions of the intraorbital repulsion $\bar{U}$,
interorbital interaction $\bar{U}'$, Hund's rule coupling $\bar{J}$, and pair hopping
energy $\bar{J}'$ in terms of a general interaction Hamiltonian
are given in Ref.~\onlinecite{ref:Kubo}
and are related to the notation in Ref.~\onlinecite{ref:GraserNJP} 
by $\bar{U}=U$, $\bar{U}'=V+J/4$, $\bar{J}=J/2$, and $\bar{J}'=J'$.
In the spin rotational invariant case the interaction parameters
are connected by $\bar{J}=\bar{J}'$ and $\bar{U}' + \bar{J}'=\bar{U}-\bar{J}$.
\begin{figure}
\includegraphics[width=.49\columnwidth]{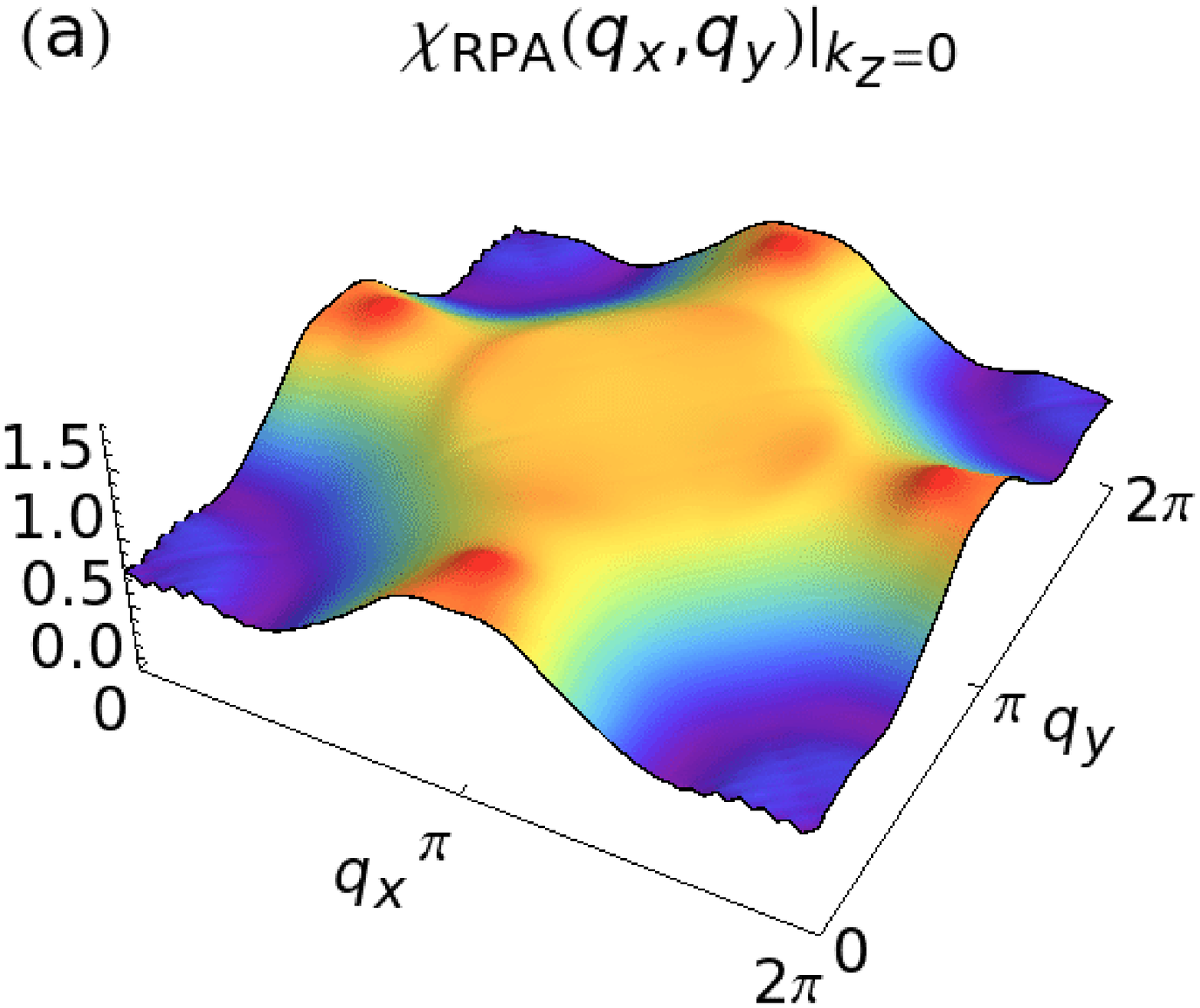}
\includegraphics[width=.49\columnwidth]{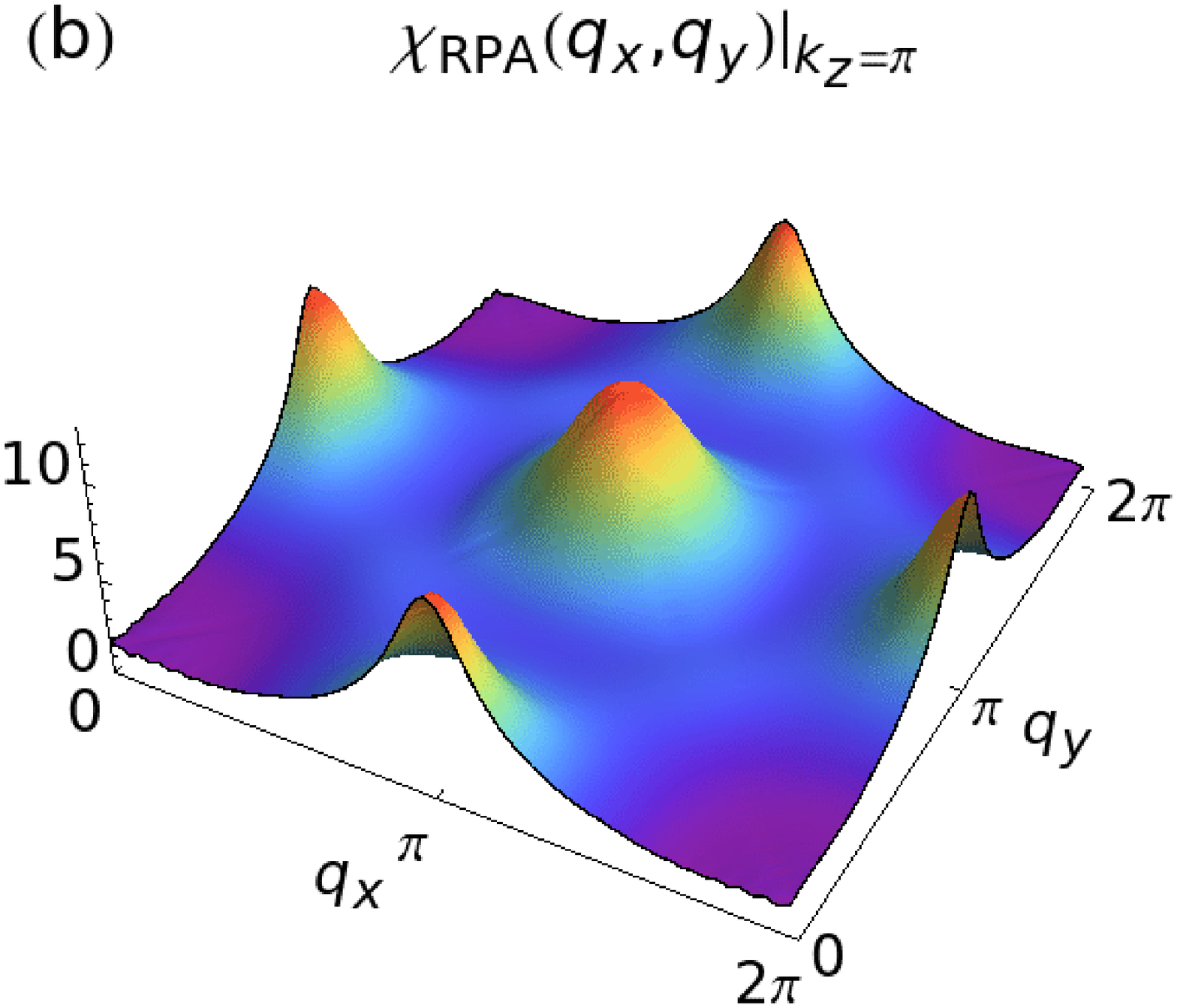}
\includegraphics[width=.49\columnwidth]{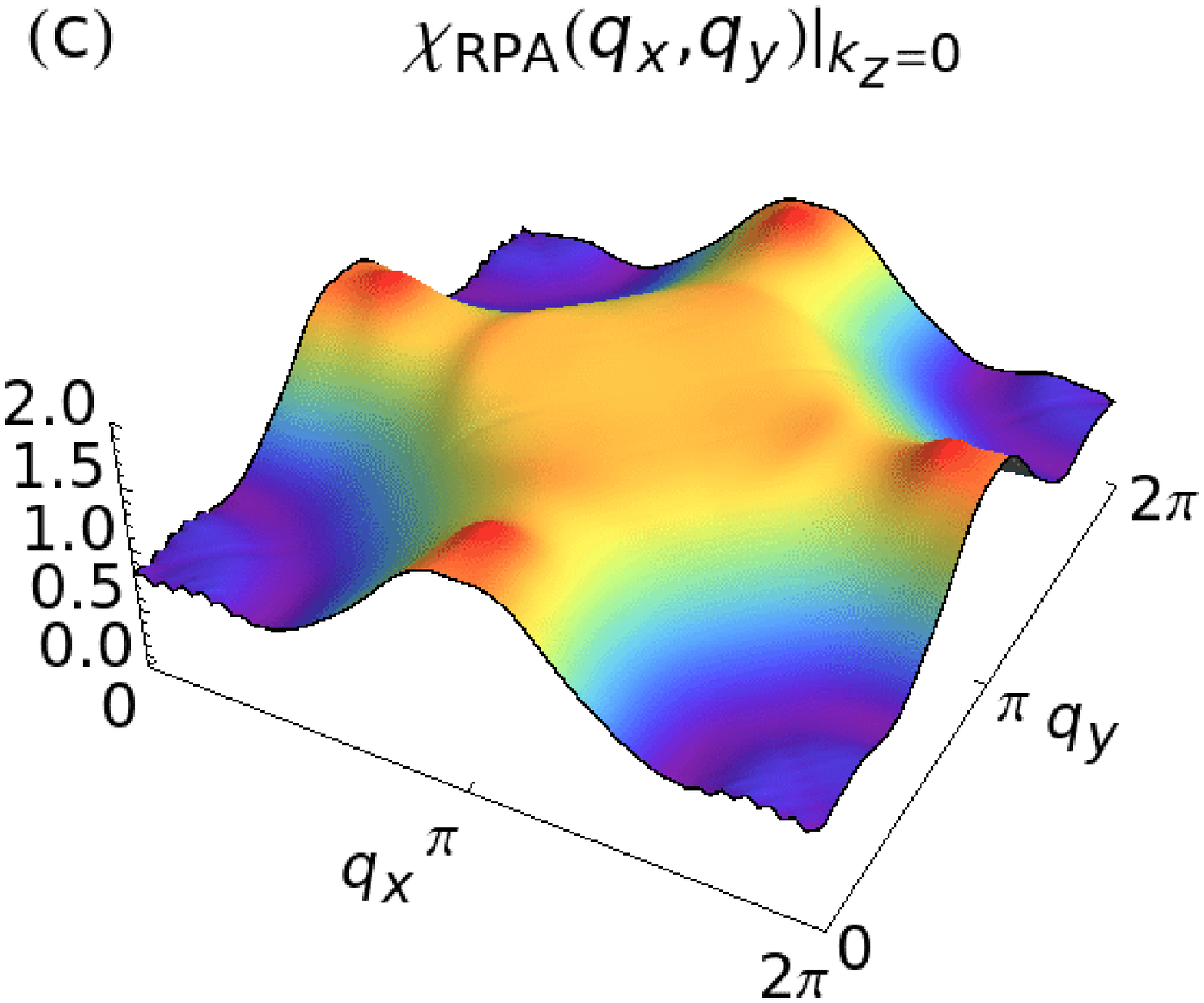}
\includegraphics[width=.49\columnwidth]{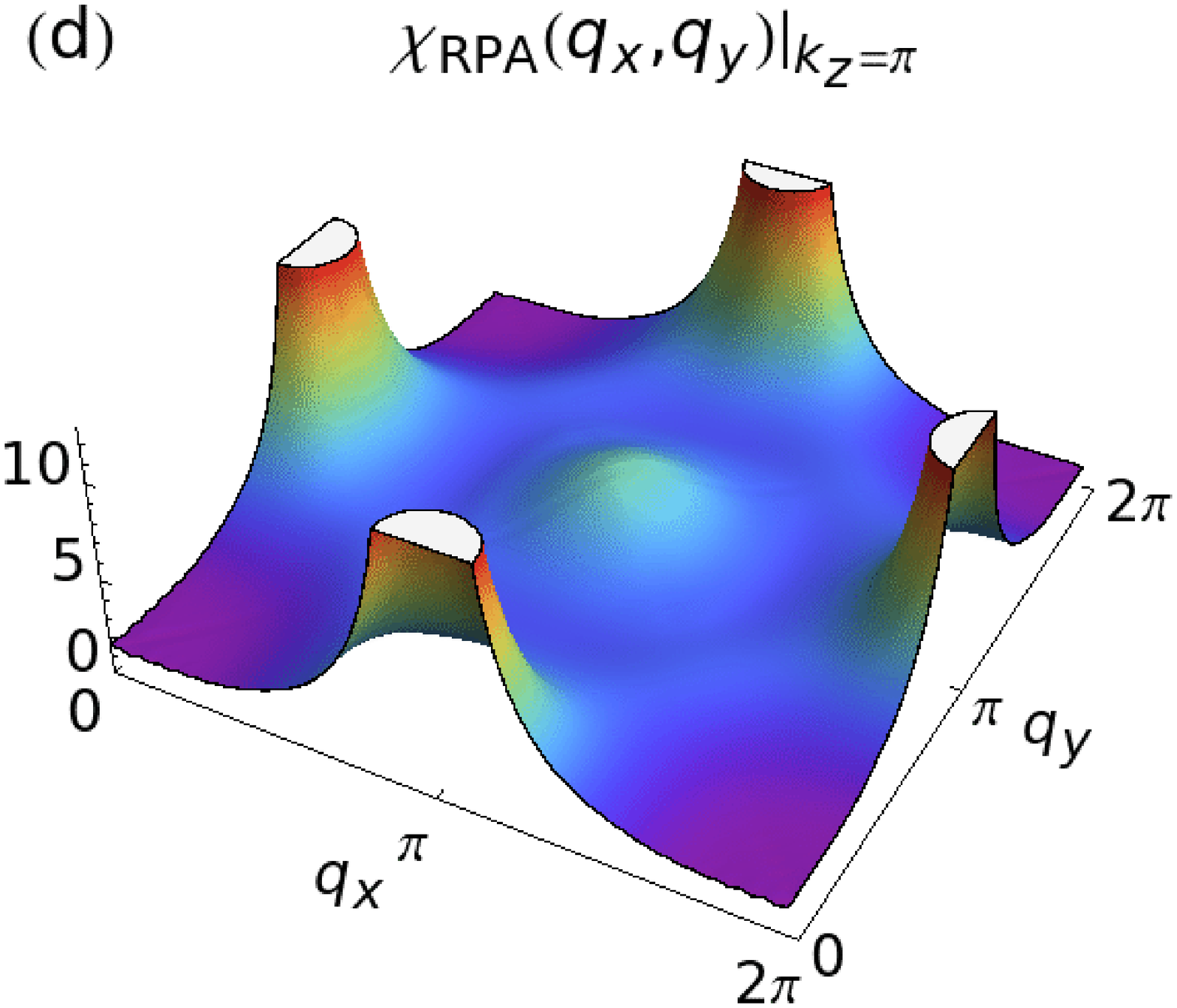}
\includegraphics[width=.4\columnwidth]{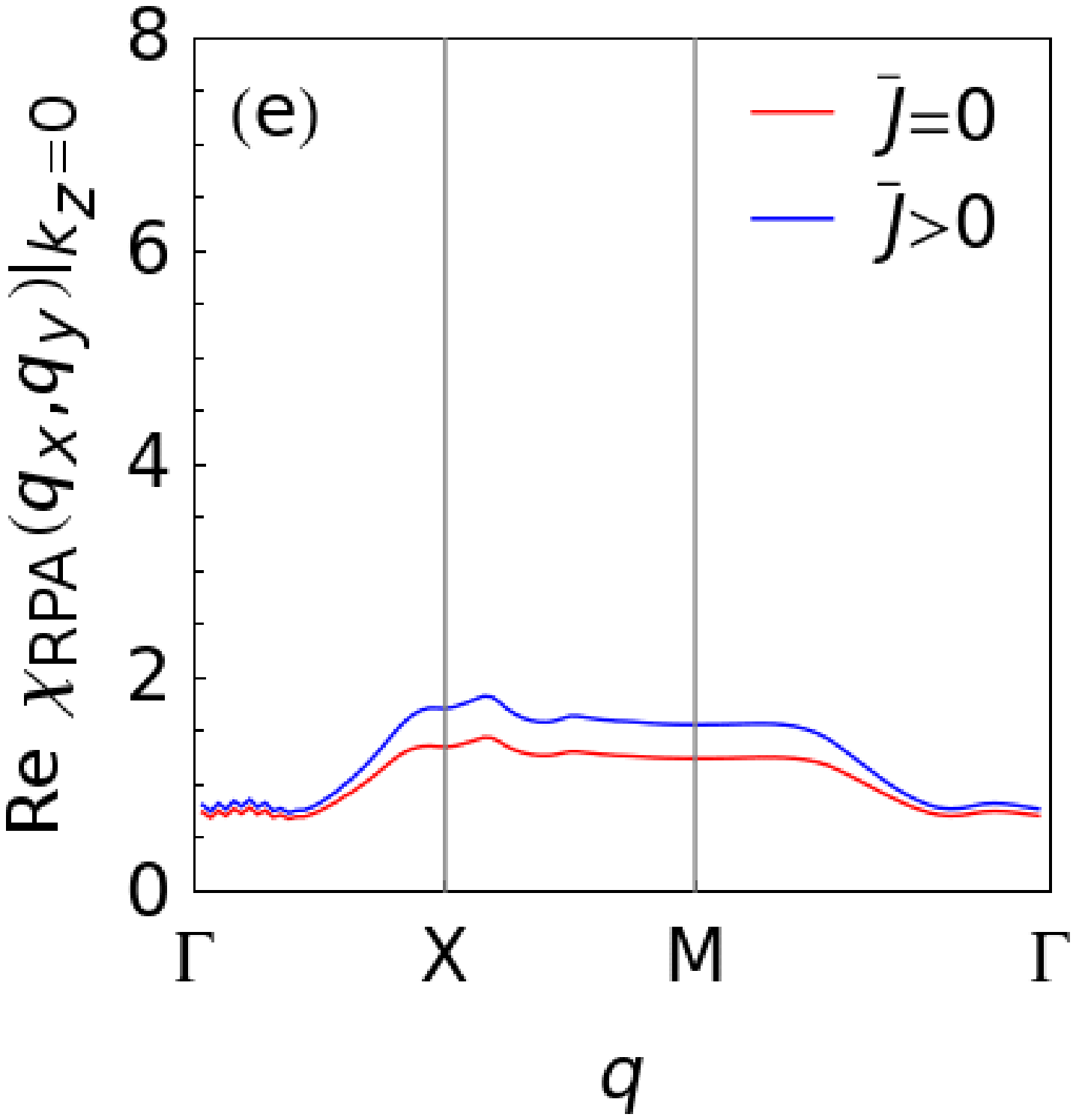} \;\;\;\;\;\;\;\;
\includegraphics[width=.4\columnwidth]{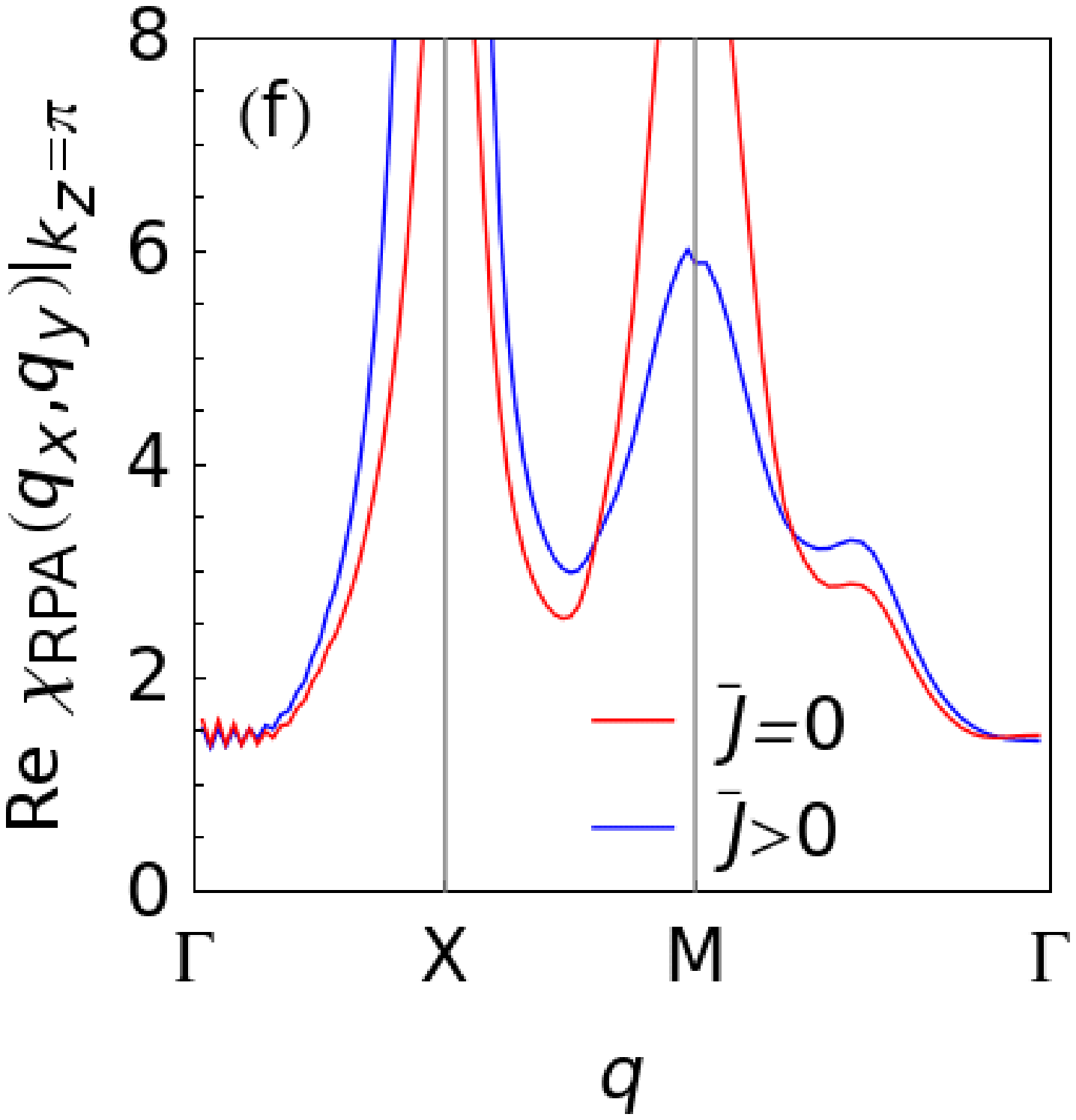}
\caption{(Color online) {\it 2D susceptibility:} The real part of the RPA
enhanced susceptibility $\chi_{\mathrm{RPA}}(q)$
as a function of the in-plane momentum transfer calculated in two dimensions
for a single value of $k_z$,
$k_z=0$ (a,c,e) and $k_z=\pi$ (b,d,f) for a hole doped compound
with $\langle n \rangle = 5.9$.
For (a) and (b) we have used $\bar{U}=0.65$ and $\bar{J}=0$, while for (c) and (d) we have used
$\bar{U}=0.55$ and $\bar{J}=0.25\bar{U}$. In panels (e) and (f) the susceptibility is shown along the
main symmetry lines with $\bar{U}=0.65$, $\bar{J}=0$ (red), and $\bar{U}=0.55$, 
$\bar{J}=0.25\bar{U}$ (blue).}
\label{fig:2Dsuscept}
\end{figure}
\begin{figure}
\includegraphics[width=.49\columnwidth]{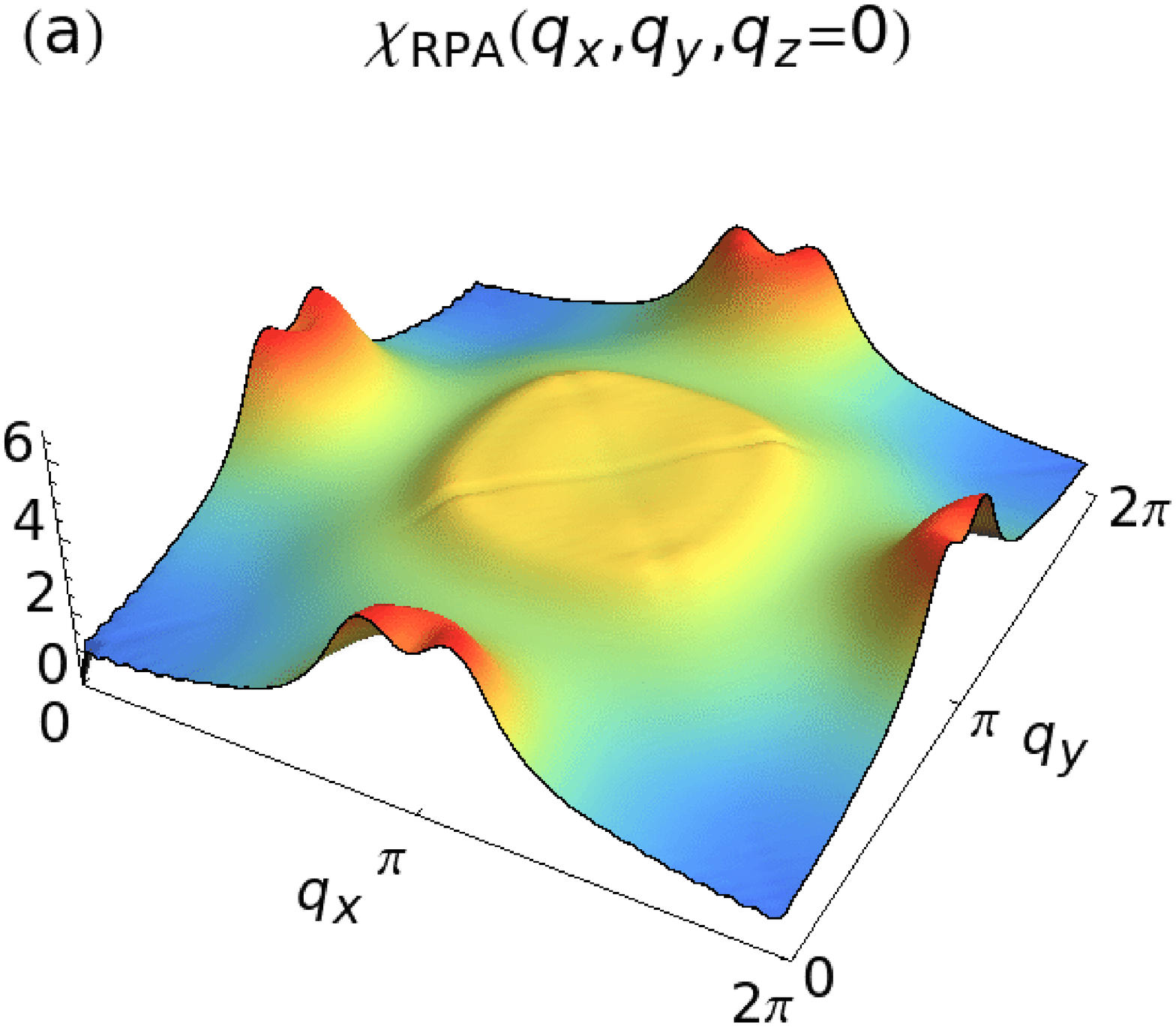}
\includegraphics[width=.49\columnwidth]{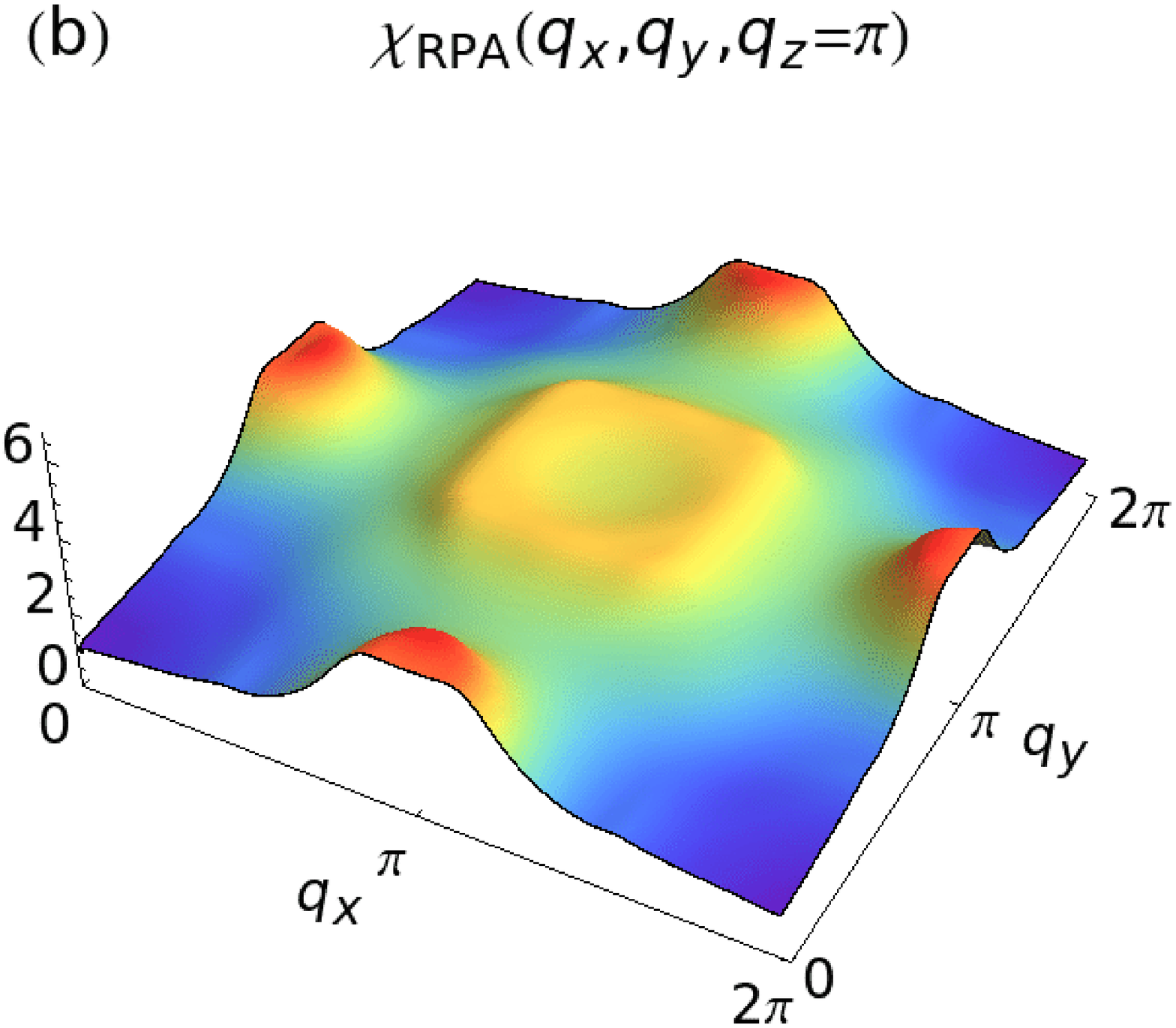}
\includegraphics[width=.49\columnwidth]{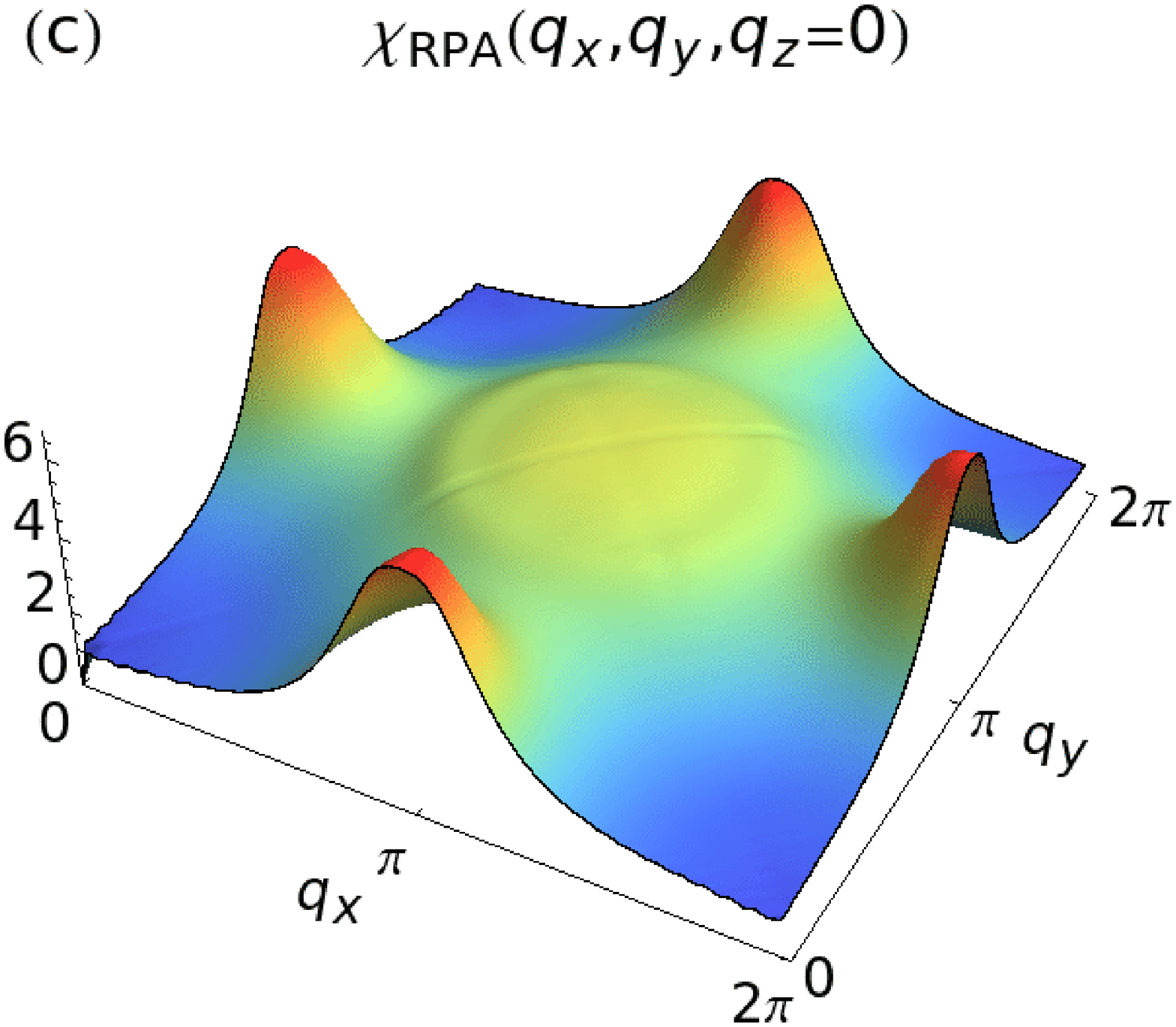}
\includegraphics[width=.49\columnwidth]{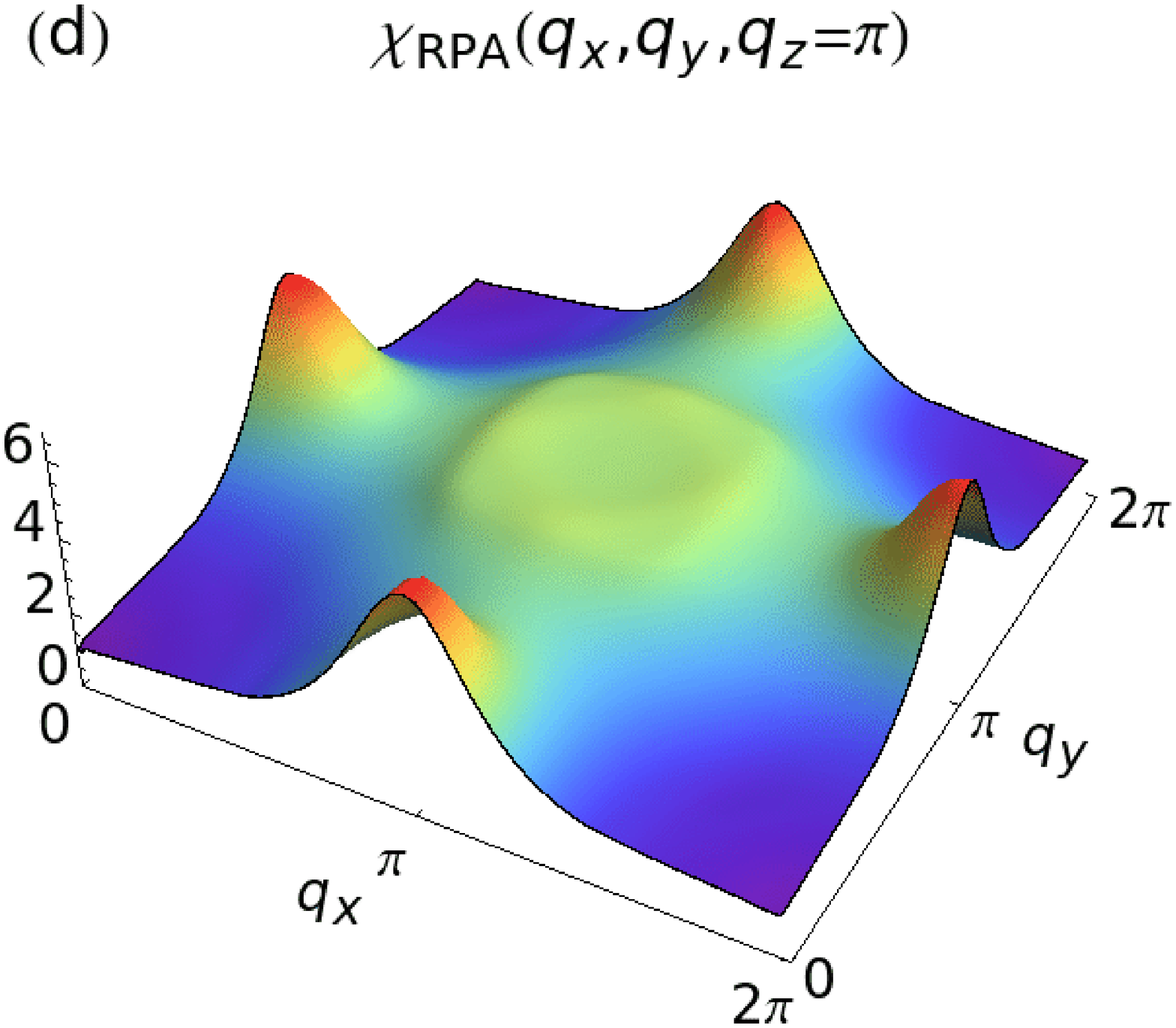}
\includegraphics[width=.4\columnwidth]{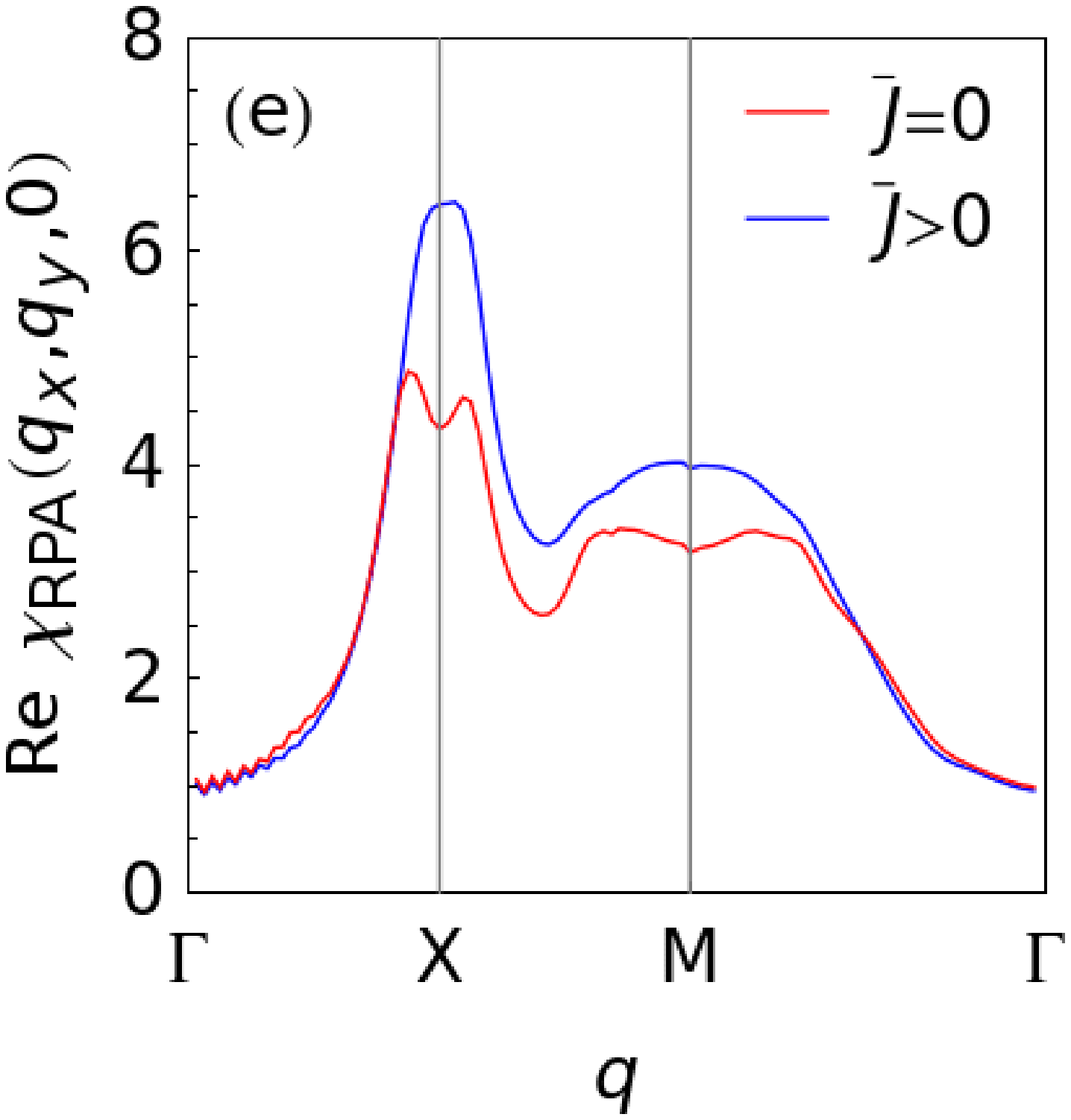} \;\;\;\;\;\;\;\;
\includegraphics[width=.4\columnwidth]{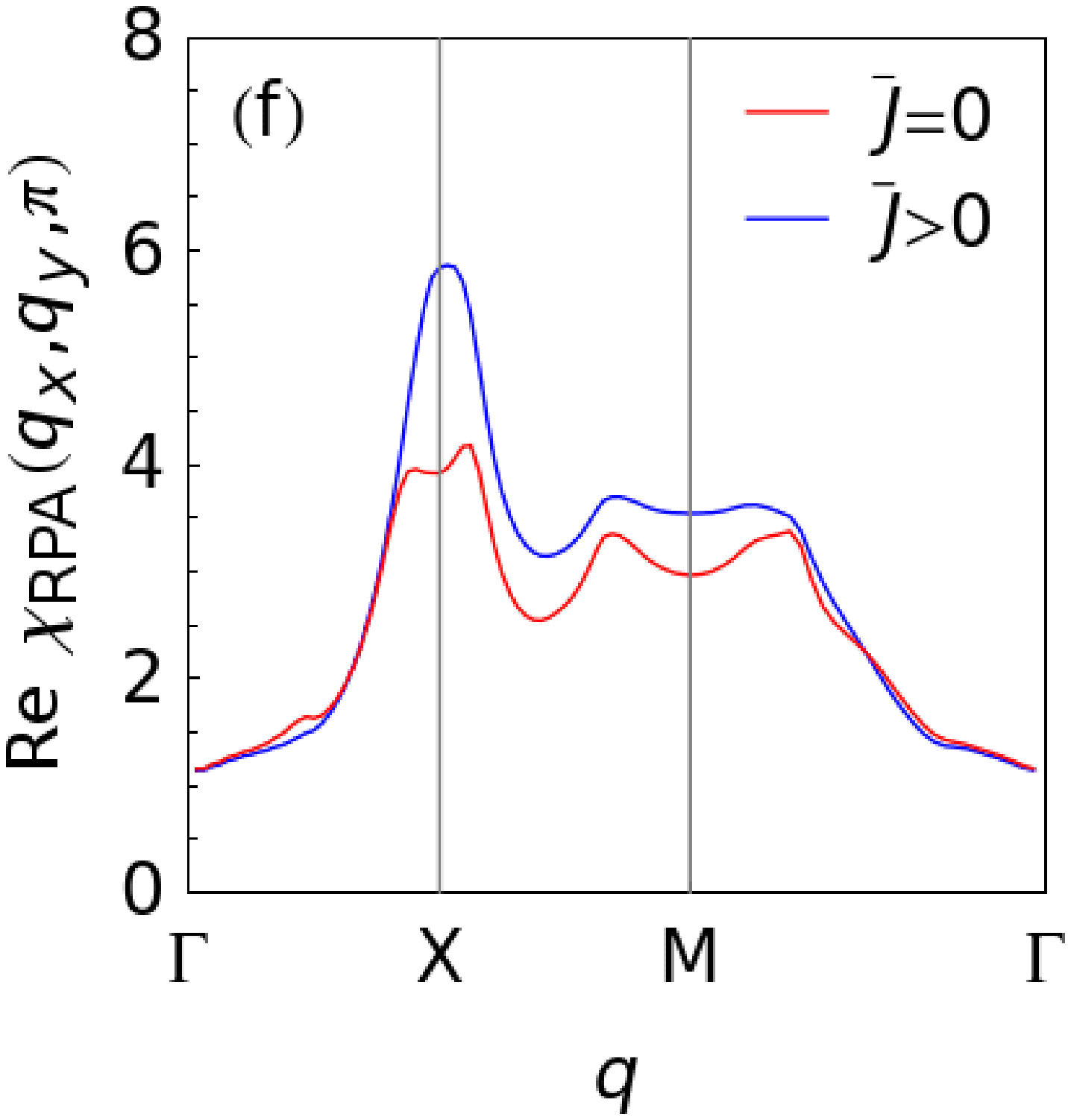}
\caption{(Color online) {\it 3D susceptibility:} The real part of the RPA
enhanced susceptibility $\chi_{\mathrm{RPA}}(q)$
as a function of the in-plane momentum transfer for two different values of $q_z$,
$q_z=0$ (a,c,e) and $q_z=\pi$ (b,d,f) for a hole doped compound
with $\langle n \rangle = 5.9$.
For (a) and (b) we have used $\bar{U}=1.1$ and $\bar{J}=0$, while for (c) and (d) we have used
$\bar{U}=0.8$ and $\bar{J}=0.25 \bar{U}$. In panels (e) and (f) we use the same coloring scheme as
in Fig.~\ref{fig:2Dsuscept}.}
\label{fig:3Dsuscept}
\end{figure}
\begin{figure}
\includegraphics[width=.49\columnwidth]{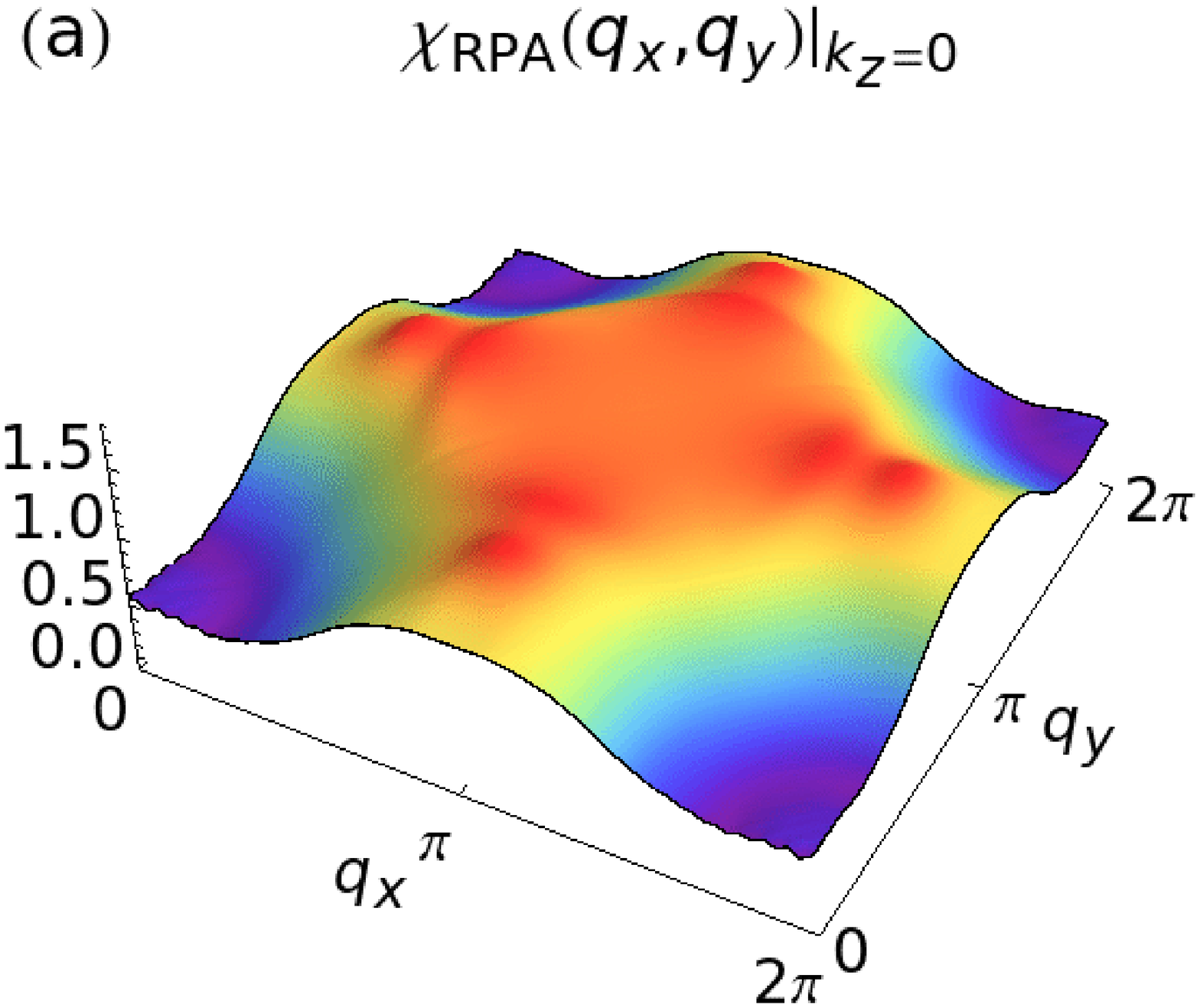}
\includegraphics[width=.49\columnwidth]{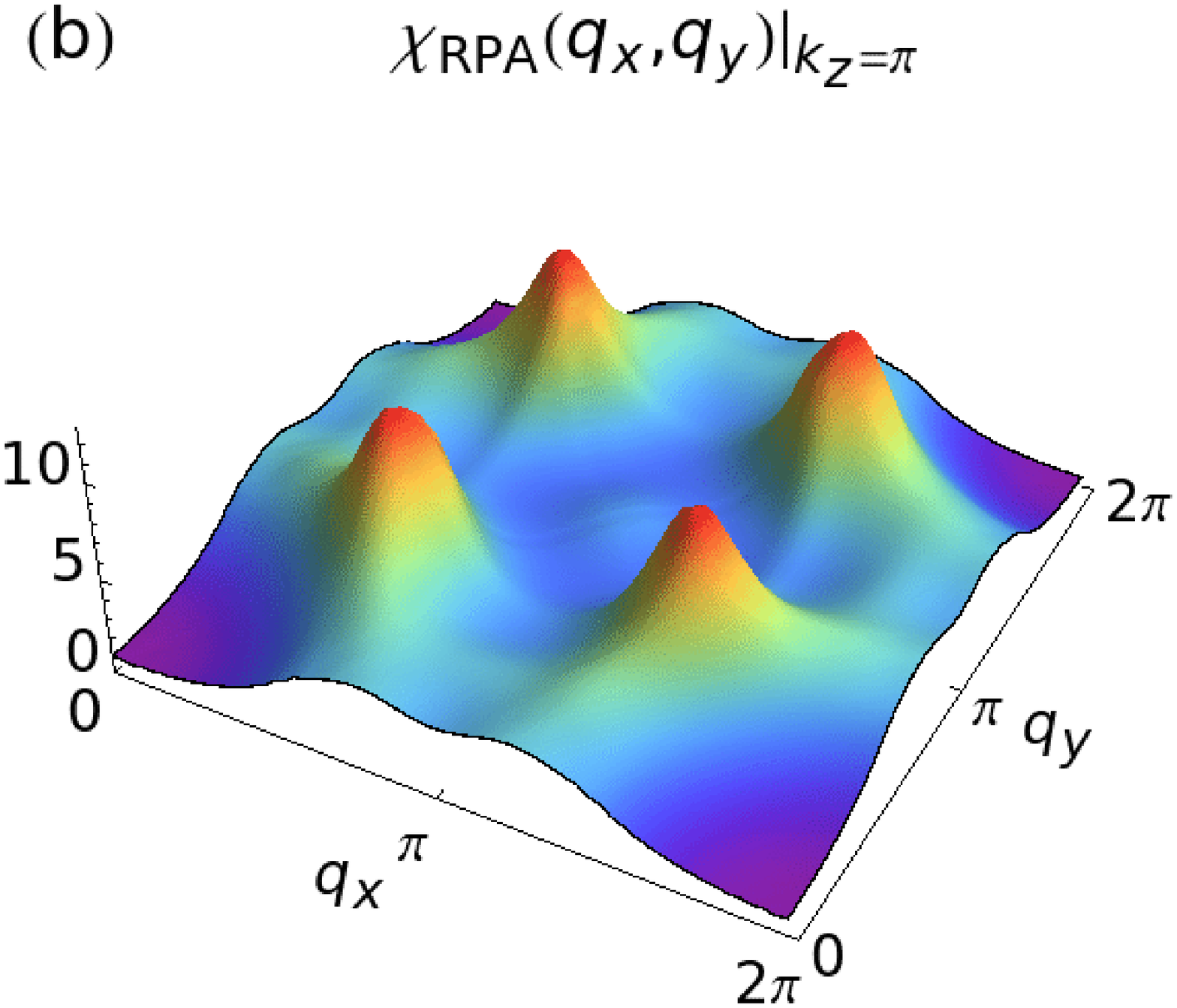}
\includegraphics[width=.49\columnwidth]{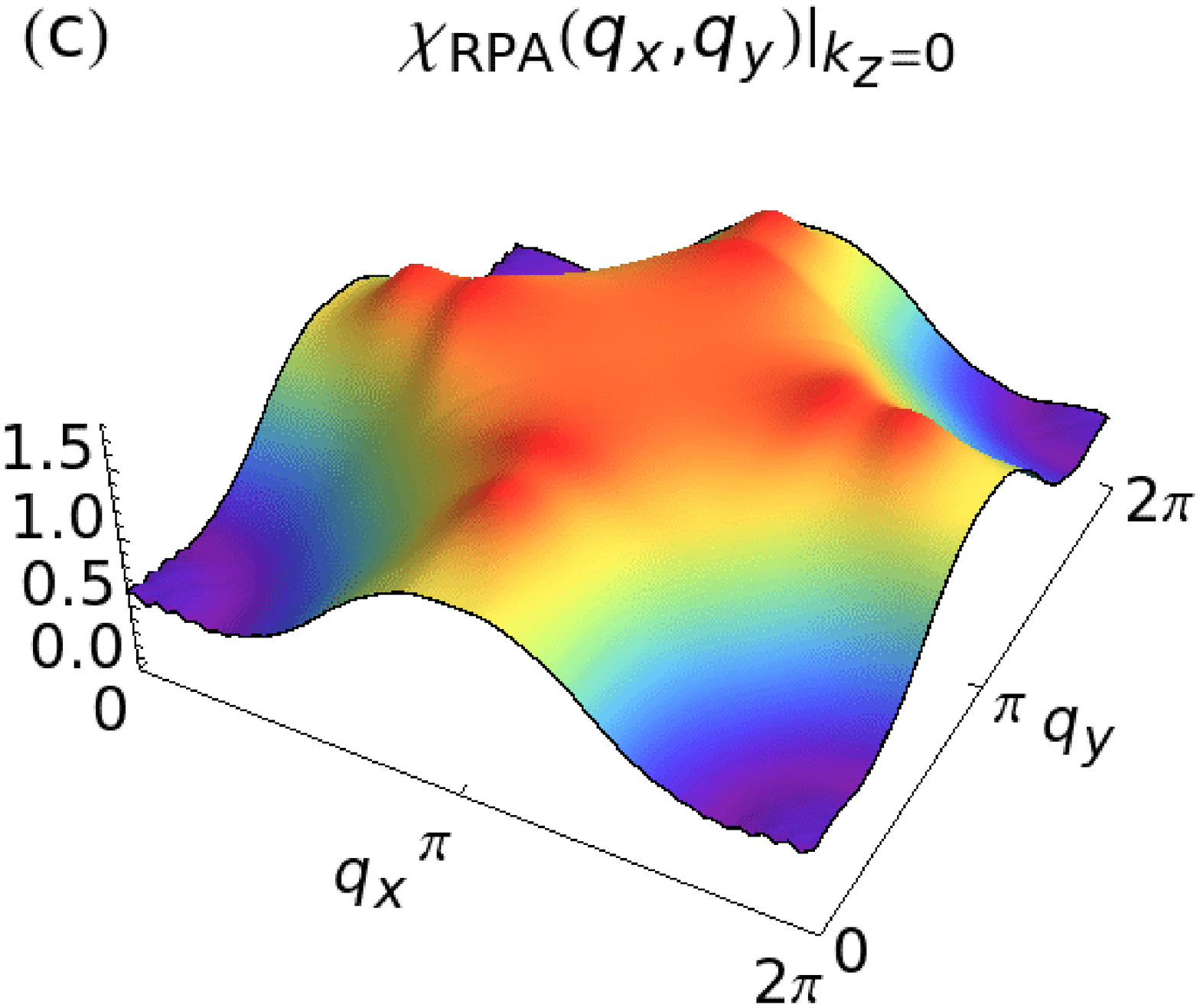}
\includegraphics[width=.49\columnwidth]{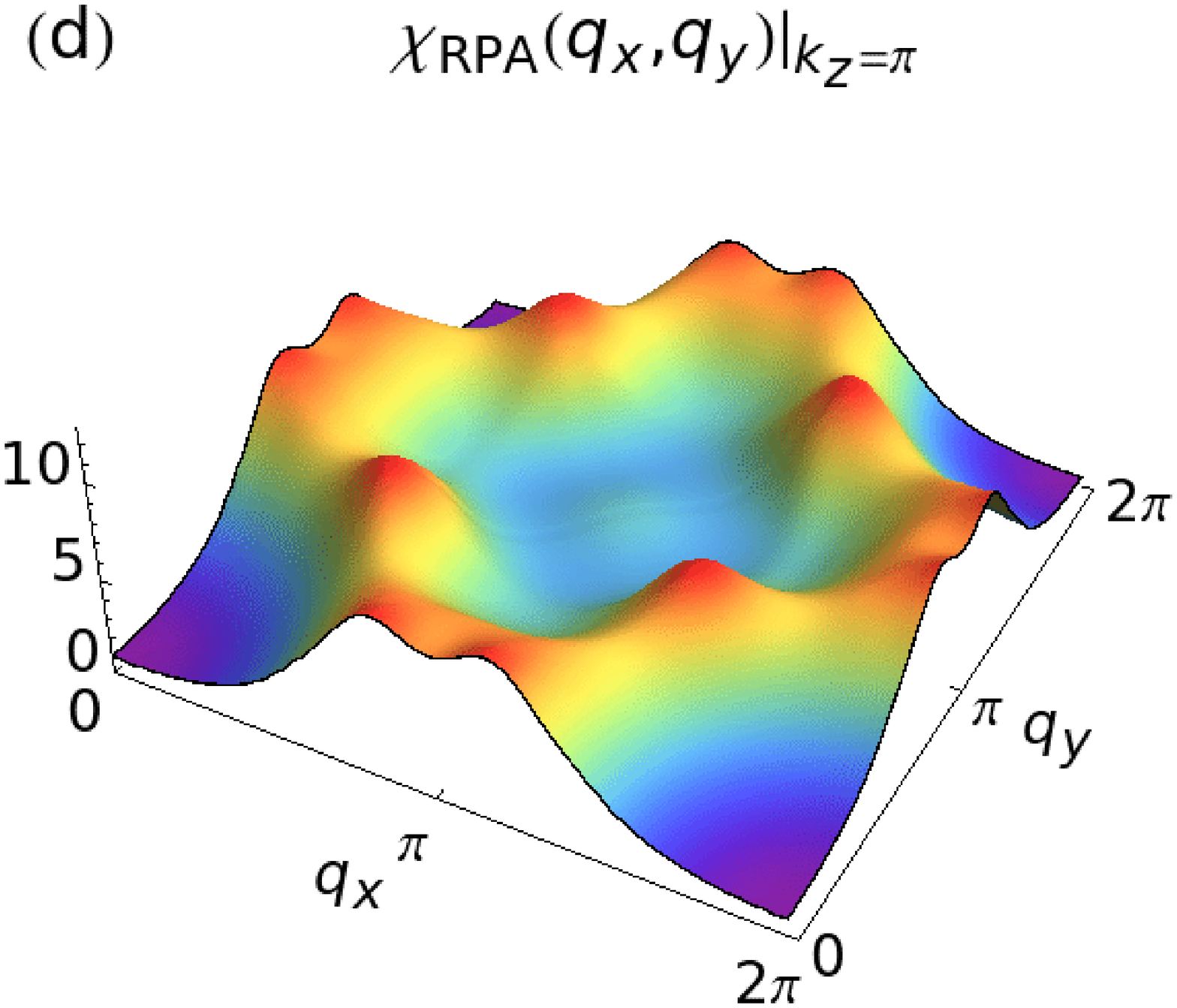}
\includegraphics[width=.4\columnwidth]{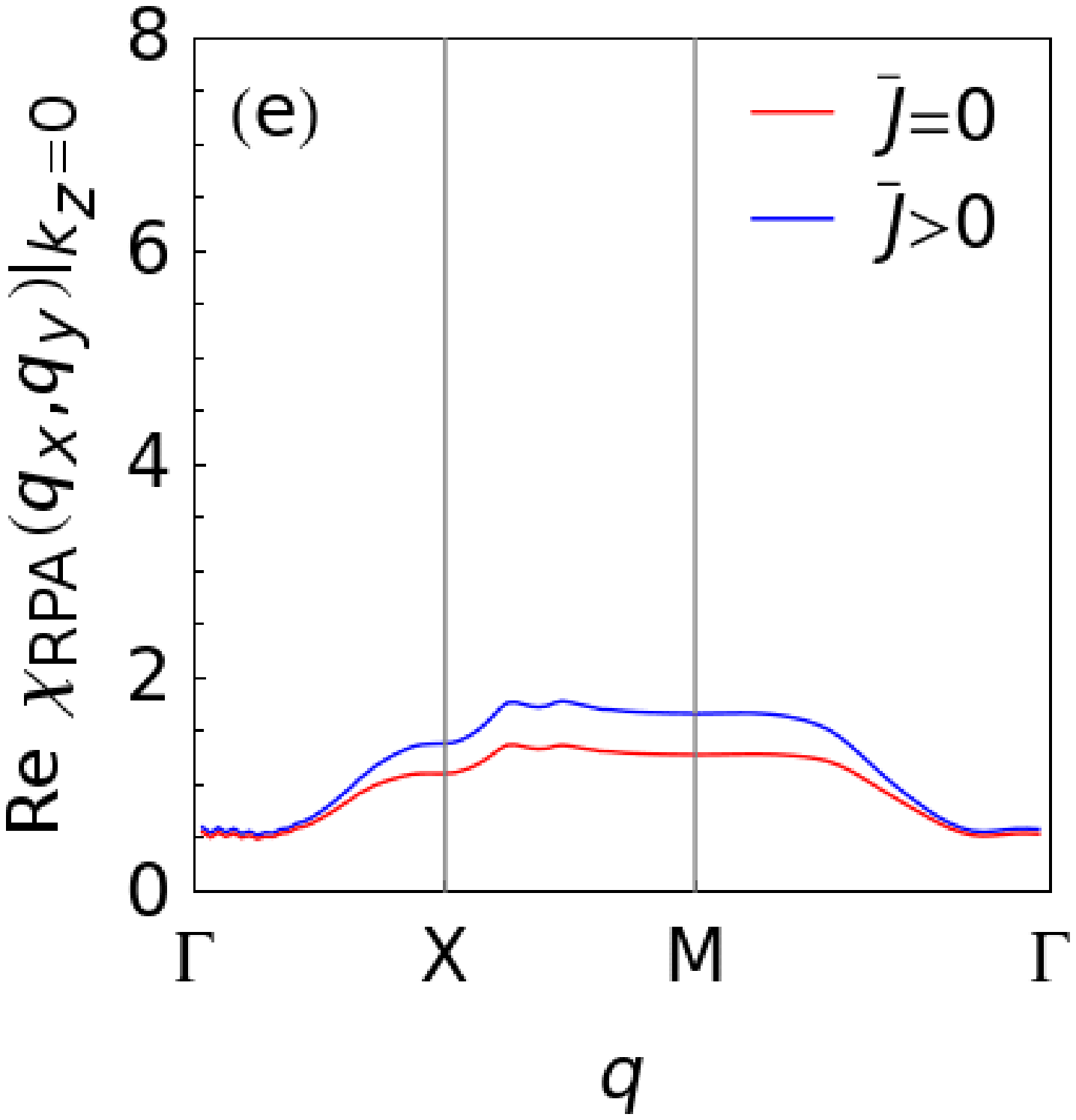} \;\;\;\;\;\;\;\;
\includegraphics[width=.4\columnwidth]{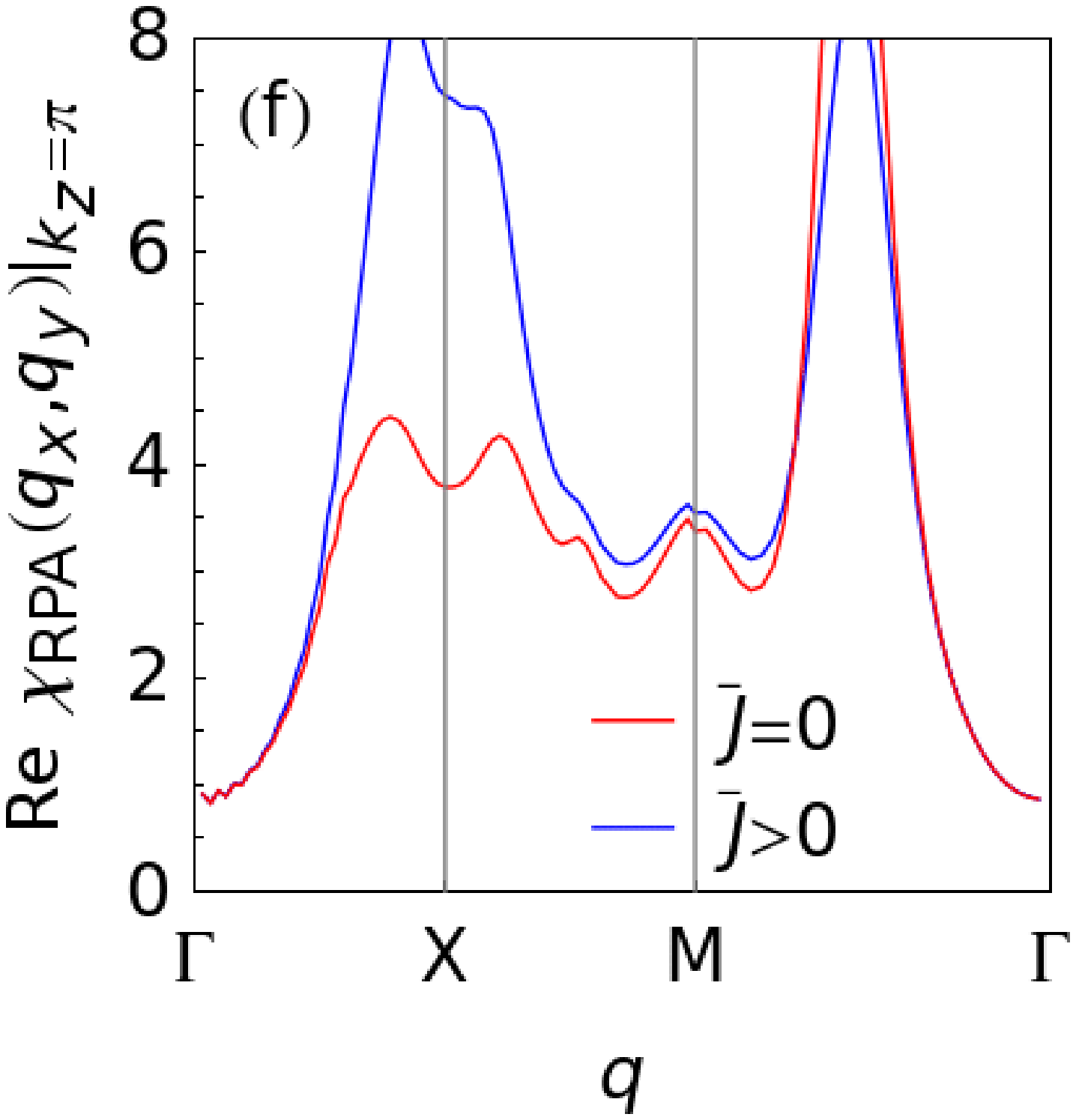}
\caption{(Color online) {\it 2D susceptibility:} The real part of the RPA
enhanced susceptibility $\chi_{\mathrm{RPA}}(q)$
as a function of the in-plane momentum transfer calculated in two dimensions
for a single value of $k_z$,
$k_z=0$ (a,c,e) and $k_z=\pi$ (b,d,f) for an undoped compound
with $\langle n \rangle = 6$.
For (a) and (b) we have used $\bar{U}=0.7$ and $\bar{J}=0$, while for (c) and (d) we have used
$\bar{U}=0.6$ and $\bar{J}=0.25 \bar{U}$. In panels (e) and (f) we use the same coloring scheme as
in Fig.~\ref{fig:2Dsuscept}.}
\label{fig:2Dsusceptx0}
\end{figure}
\begin{figure}
\includegraphics[width=.49\columnwidth]{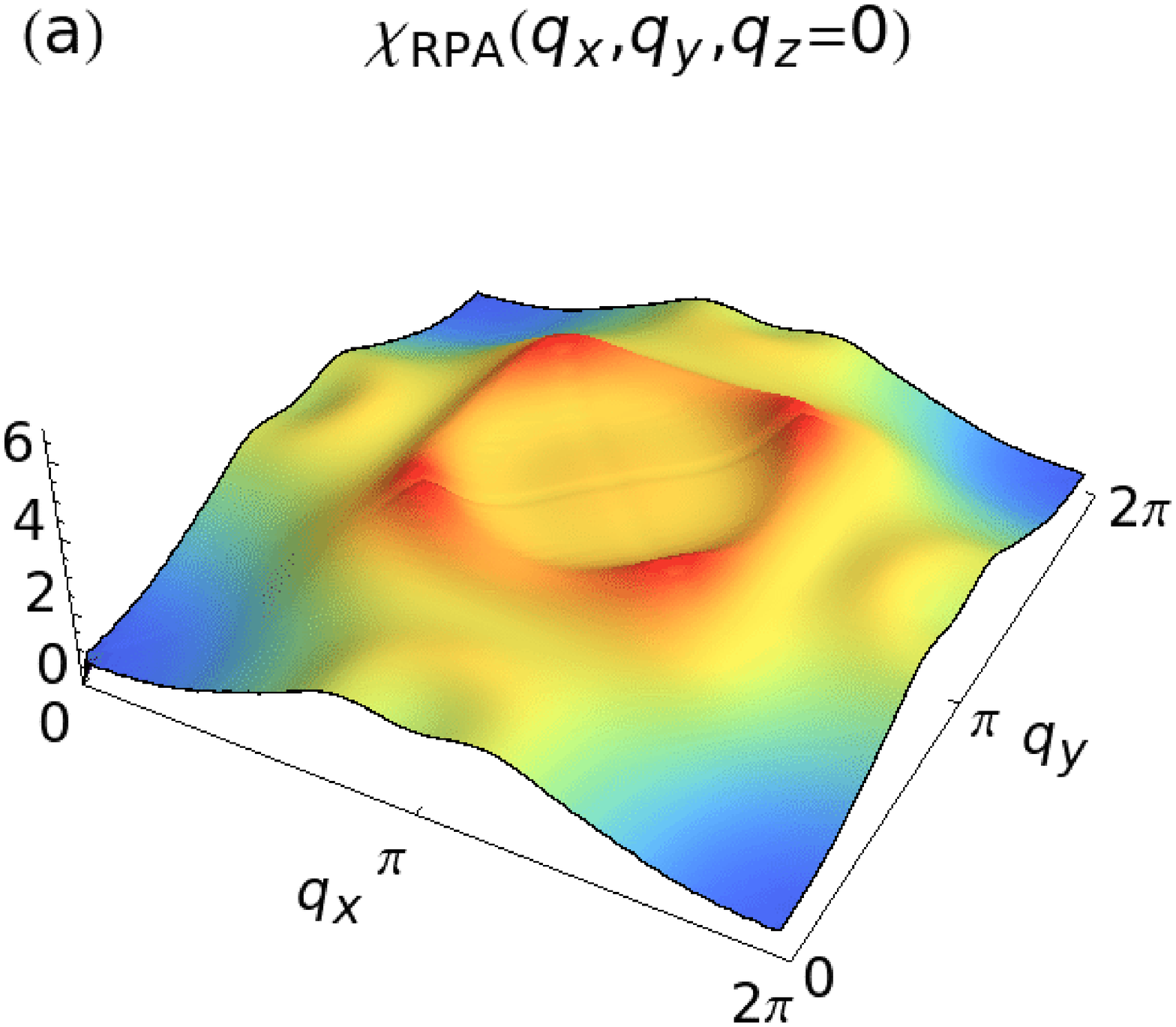}
\includegraphics[width=.49\columnwidth]{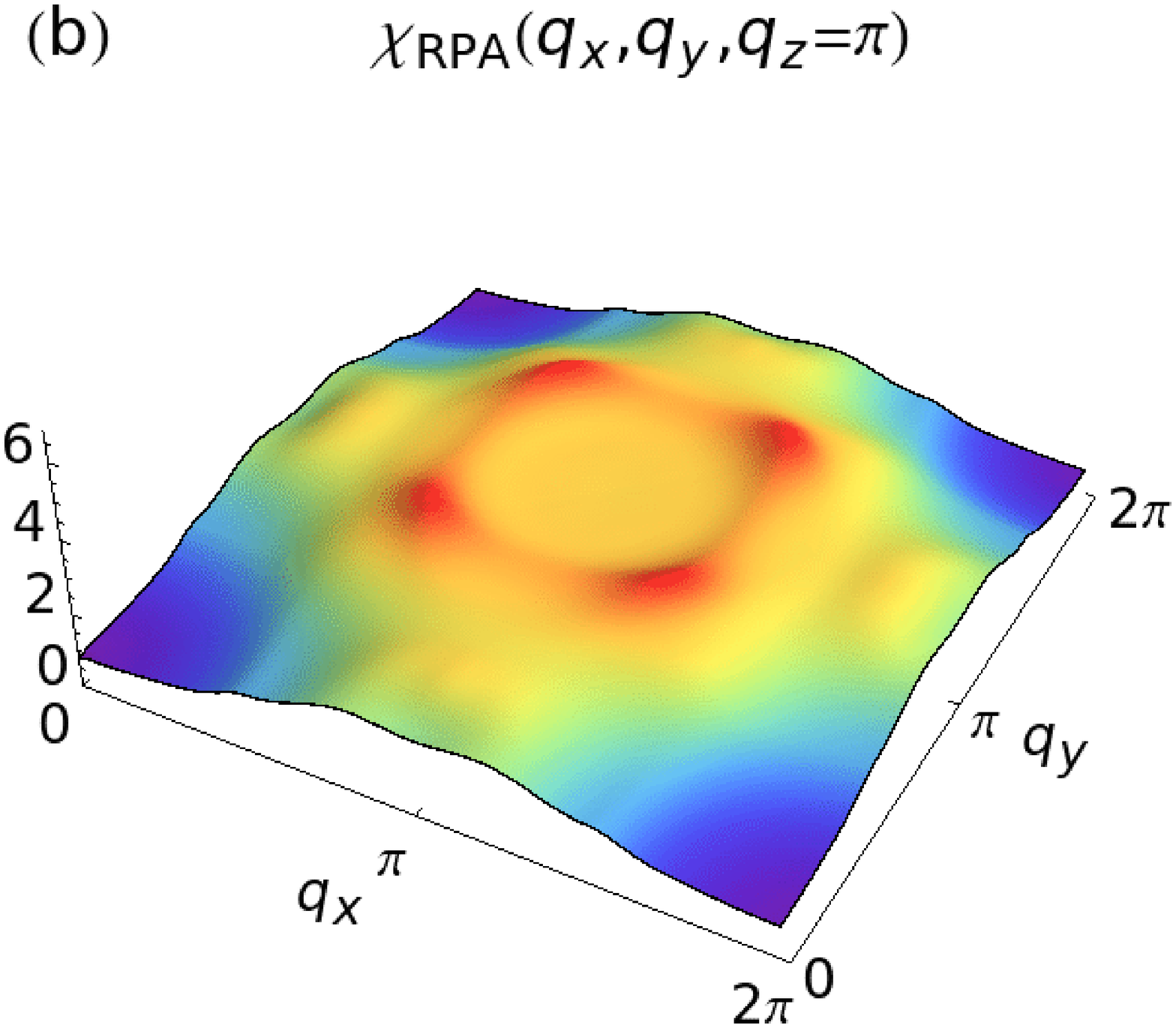}
\includegraphics[width=.49\columnwidth]{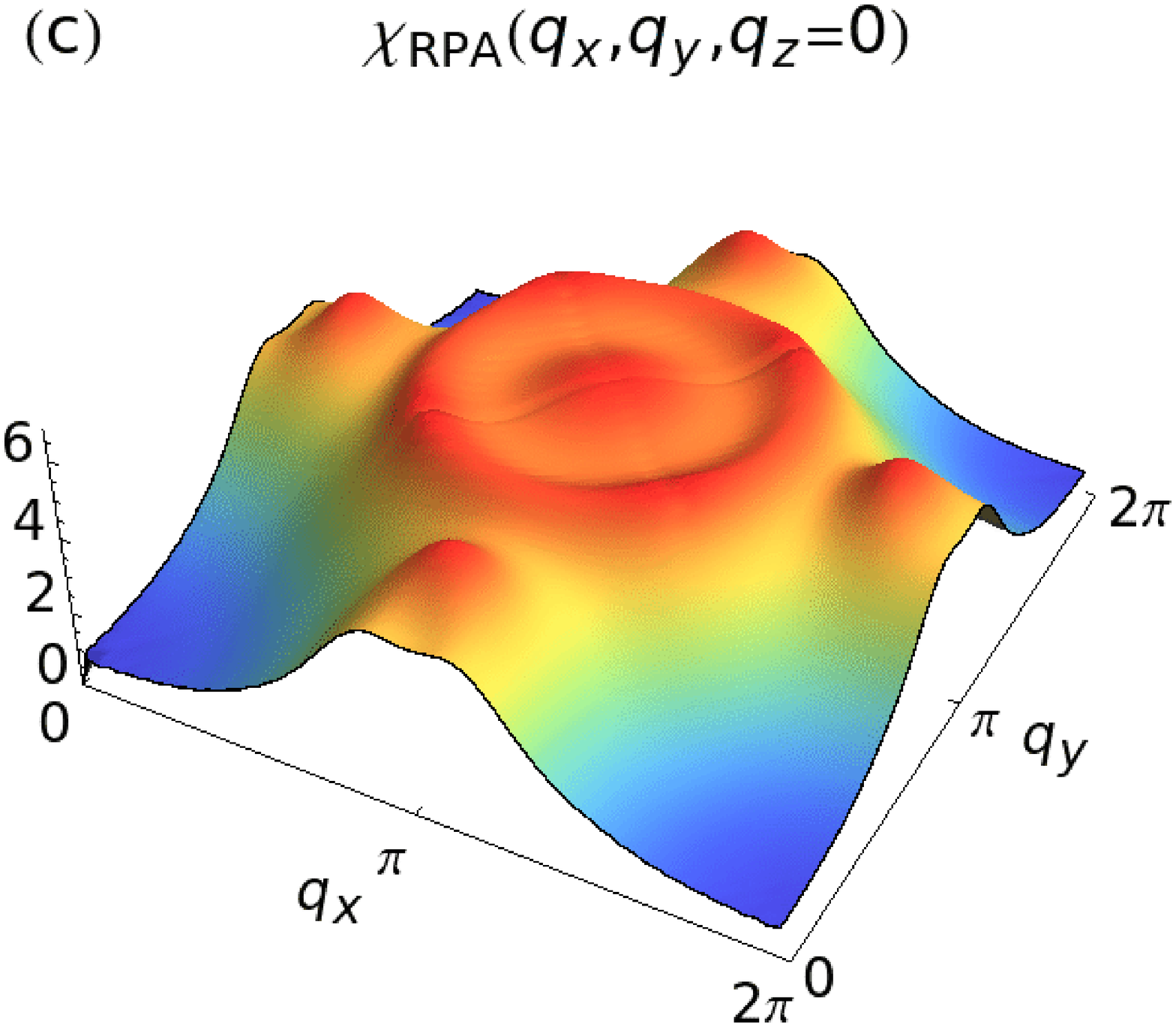}
\includegraphics[width=.49\columnwidth]{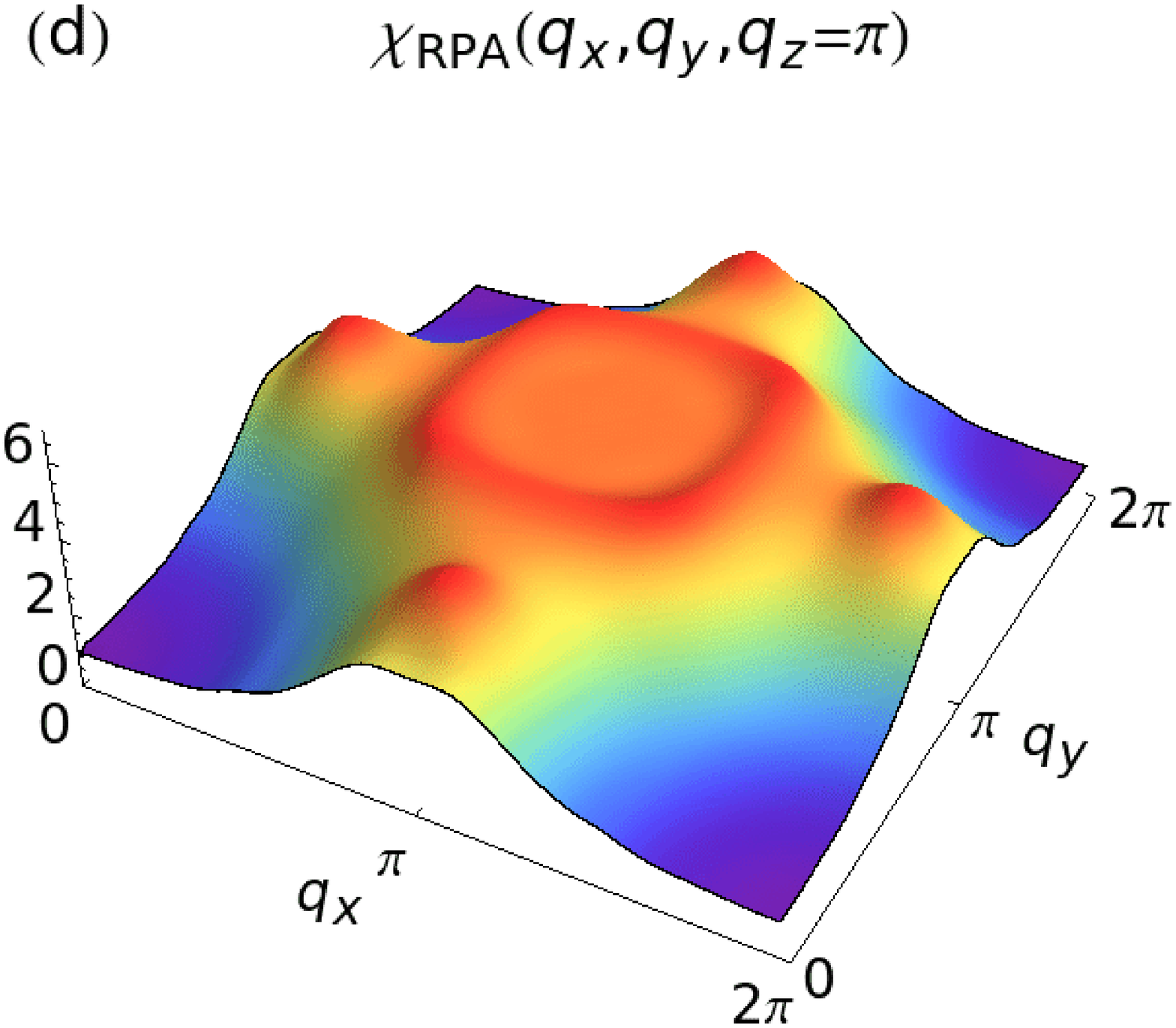}
\includegraphics[width=.4\columnwidth]{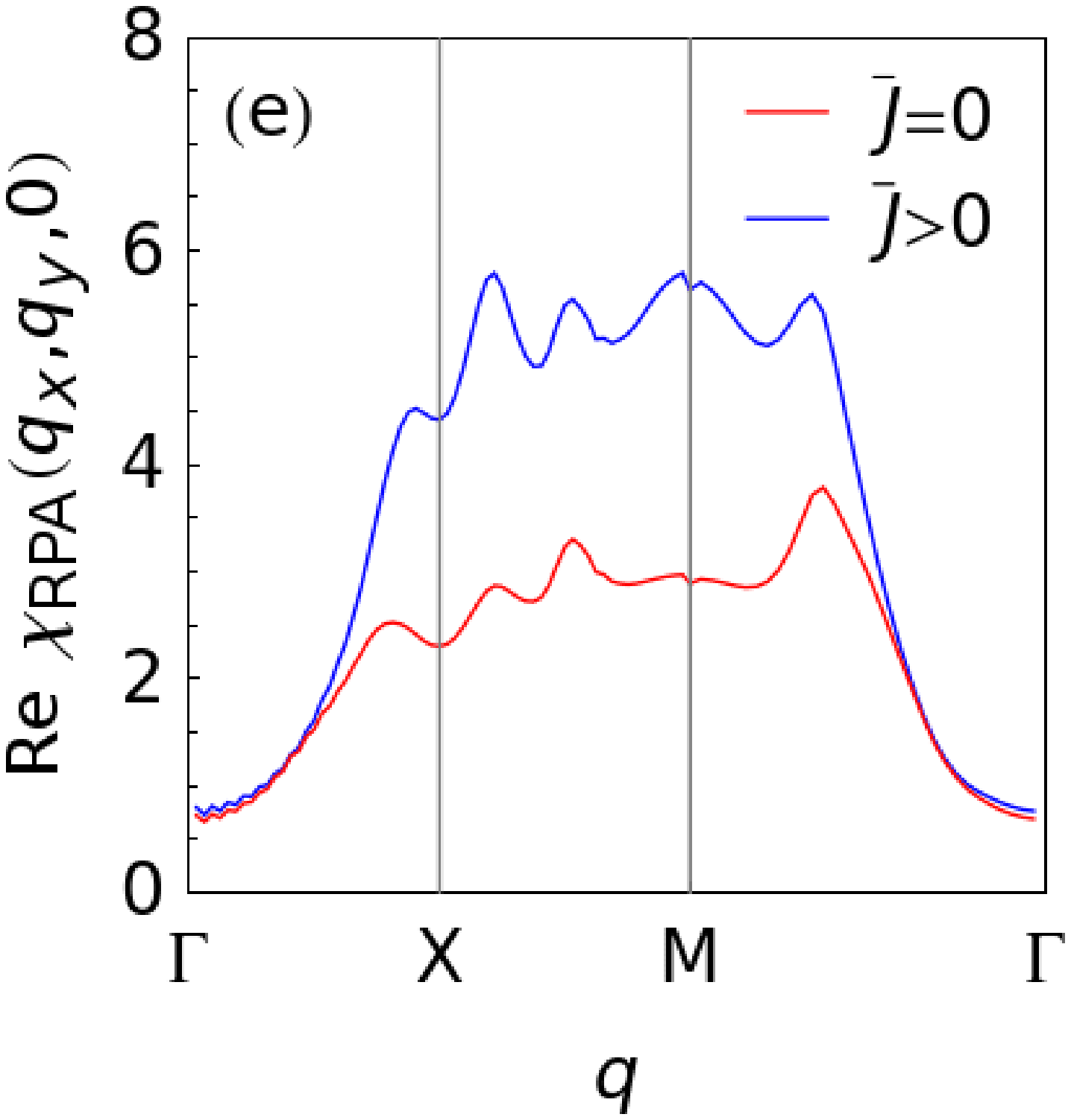} \;\;\;\;\;\;\;\;
\includegraphics[width=.4\columnwidth]{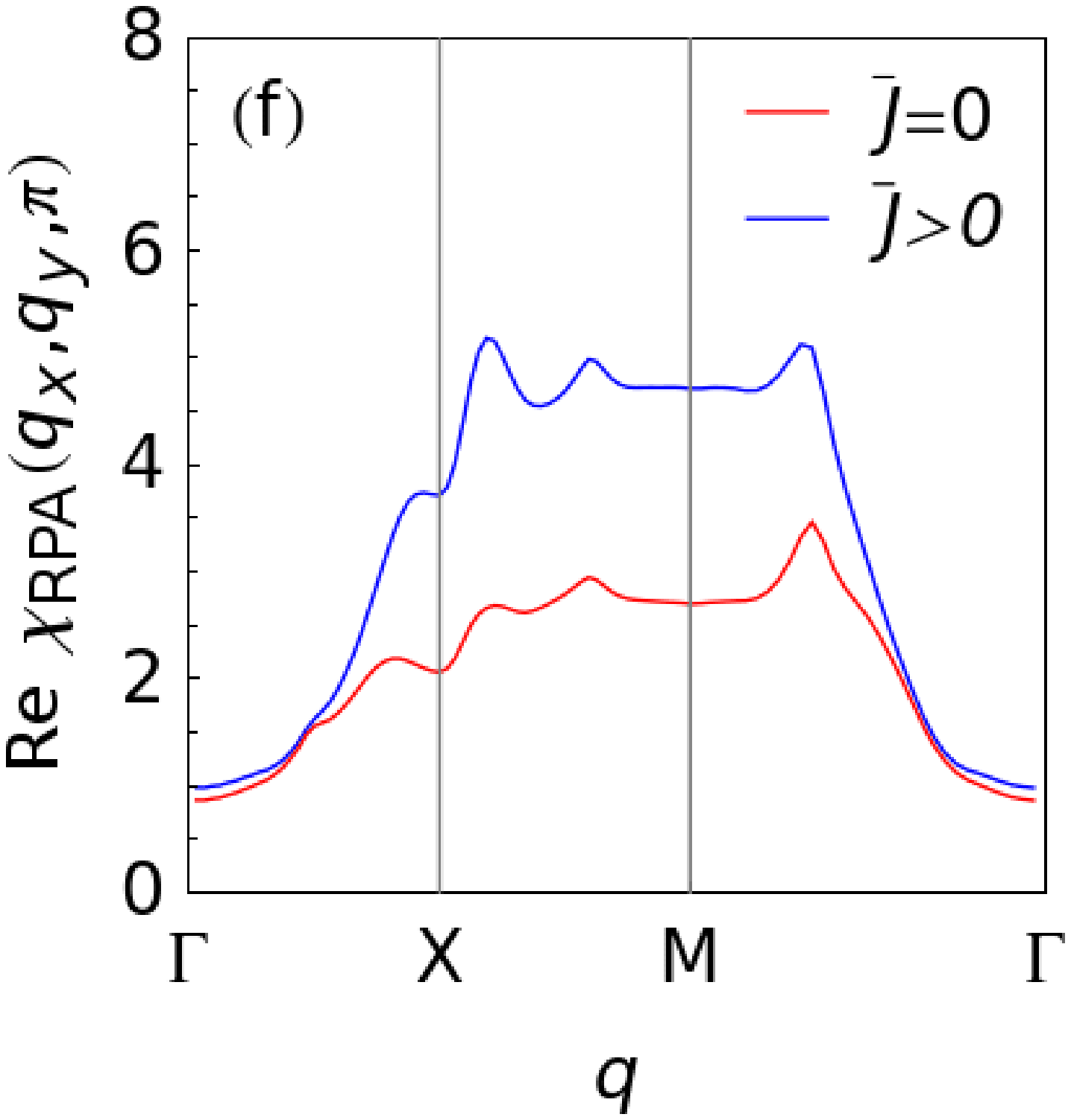}
\caption{(Color online) {\it 3D susceptibility:} The real part of
the RPA enhanced susceptibility $\chi_{\mathrm{RPA}}(q)$
as a function of the in-plane momentum transfer for two different values of $q_z$,
$q_z=0$ (a,c,e) and $q_z=\pi$ (b,d,f) for an undoped compound
with $\langle n \rangle = 6$.
For (a) and (b) we have used $\bar{U}=1.1$ and $\bar{J}=0$, while for (c) and (d) we have used
$\bar{U}=0.9$ and $\bar{J}=0.25 \bar{U}$. In panels (e) and (f) we use the same coloring scheme as
in Fig.~\ref{fig:2Dsuscept}.}
\label{fig:3Dsusceptx0}
\end{figure}
The charge susceptibility $(\chi_0^{\rm RPA})_{stpq}$ can be derived
similarly with a different interaction matrix $U^c$ with the 
components
\begin{equation}
U_{aaaa}^c = \bar{U}, \, U_{bbaa}^c = 2\bar{U}'-\bar{J}, \, 
U_{abab}^c = 2\bar{J}-\bar{U}', \, U_{abba}^c = \bar{J}'. \nonumber
\end{equation}
In the following we will distinguish between sets of interaction
parameters with and without finite Hund's rule coupling
(and corresponding pair hopping) and we will adjust the interaction parameters
to be close to the superconducting instability.
Note that the ratio of $\bar{U}/\bar{J}$ taken here for 
$\bar{J}>0$ cases is similar to that found by Miyake {\it et al.}~\cite{ref:Miyake} 
from {\it ab initio} calculations, but that the overall scale is smaller.  
This renormalized RPA interaction scale is familiar from one-band interacting 
models in the cuprates~\cite{ref:BulutScalapino}.
In Fig.~\ref{fig:2Dsuscept}, we show the real part of the 
RPA enhanced spin susceptibility
\begin{equation}
\chi_S({\bf q})=\frac{1}{2} \sum_{sp} (\chi_1)_{sspp}({\bf q},0)
\end{equation}
calculated over a 2D Fermi surface corresponding to a cut through the
3D Fermi surface at fixed  $k_z$=0 or $\pi$, and for two different
sets of interaction parameters. 
Here we find that
the susceptibility calculated at $k_z=0$ shows only a small incommensurate
enhancement of the main scattering peaks at ${\bf q}=(\pi,0)$, while the
susceptibility calculated at $k_z=\pi$ is commensurate and strongly peaked for the same
value of the momentum transfer.
This enhancement at $k_z=\pi$ can be explained by the additional intraorbital
scattering channel between the electron and hole pockets
due to a major $d_{xy}$ contribution 
on the $\alpha_1$ FS sheet at $k_z=\pi$ that is absent at 
$k_z=0$ (see Fig.~\ref{fig:FS}).
Comparing the susceptibility along the main symmetry lines in
panel (e) and (f), we find that for the $k_z = \pi$ cut the
non-zero Hund's rule coupling $\bar{J}=0.25 \bar{U}$ 
together with the related pair hopping
$\bar{J}'=\bar{J}$ leads to a strong enhancement of the 
scattering peak at ${\bf q} = (\pi,0)$ exceeding the 
scattering peak at ${\bf q}=(\pi,\pi)$. 
By contrast, an exclusive increase of the 
intra-orbital interaction $\bar{U}$ without Hund's rule coupling
$\bar{J}=0$, chosen to produce a comparable 
enhancement of the scattering peak at ${\bf q}=(\pi,0)$, 
also enhances the scattering peak at ${\bf q}=(\pi,\pi)$.

Comparing the results for the two different $k_z$ values
we therefore expect that for the full 3D calculation, contributions from
$k_z=\pi$ will dominate the total susceptibility but the structures will be
less sharp as a result of the $k_z$ averaging process.
In Fig.~\ref{fig:3Dsuscept}, we have now studied the full 3D susceptibility
including the explicit $q_z$ dependence and summing over $k_z$. As expected
from our two dimensional study the main peak structure is similar to the
dominating $k_z=\pi$ cut of the susceptibility. But we note that for $\bar{J}=0$
the well separated scattering peak at ${\bf q} = (\pi,\pi)$
in the $k_z = \pi$ susceptibility is already broadened to a plateau-like
structure with its main weight shifted to an incommensurate position around
${\bf q}=(\pi,\pi)$. For $\bar{J}>0$ the substructures in the scattering peaks vanish
and the susceptibility shows broad but well developed 
$(\pi,0)$ scattering peaks at a commensurate position.
Note this result is quite similar to the commensurate normal state neutron scattering 
intensity observed by Inosov {\it et al.}~\cite{ref:Inosov}, which was difficult to understand 
in the 2D calculations for the 1111 Fermi surface in Ref.~\onlinecite{ref:GraserNJP}.  
Here we find that the peaks in Re $\chi({\bf q},\omega=0)$ tend to correspond to 
those of Im $\chi({\bf q}, \omega)$ for small $\omega$, suggesting that the 
``averaging'' of the susceptibility due to the 3D dispersion 
in the 122 materials is sufficient to account for the commensurate response.

In Figs.~\ref{fig:2Dsusceptx0} and \ref{fig:3Dsusceptx0}, we show the equivalent results 
to Figs.~\ref{fig:2Dsuscept} and \ref{fig:3Dsuscept} but for zero doping.  
Compared to the hole doped case, we see that the peaks at $(\pi,0)$ are 
considerably suppressed (see Fig.~\ref{fig:2Dsusceptx0}), and new incommensurate peaks along 
the $(\pi,0)$-$(0,\pi)$ line appear. In the fully integrated 3D result, 
Fig.~\ref{fig:3Dsusceptx0}, by contrast, the nearly commensurate response at 
$(\pi,0)$ is recovered, but is suppressed relative to an incommensurate ridge 
of response.  
  
\section{Pairing symmetry}
\label{sec:pair}

\begin{figure}
\includegraphics[width=\columnwidth]{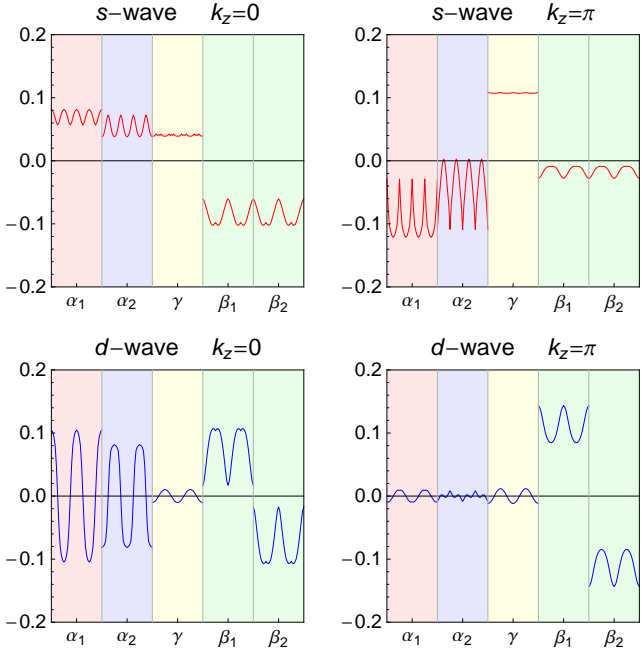}
\caption{(Color online) {\it 2D pairing functions, $\bar{J}=0$:}  The leading (upper row) 
and subleading (lower row) pairing function 
for the hole doped compound ($\langle n \rangle = 5.9$) 
plotted along the Fermi surfaces at two different $k_z$ cuts. 
The pairing functions are shown in the order $\alpha_1$, $\alpha_2$, 
$\gamma$, $\beta_1$, and $\beta_2$ running counter-clockwise around each Fermi surface sheet
with the rightmost point as the starting point on each sheet, except the $\beta_2$
pocket where the plots start with the uppermost point.
The calculations were performed for $\bar{U} = 0.65$ and $\bar{J}= 0$ and the eigenvalues are 
$\lambda=0.022$ ($s$-wave) and $\lambda=0.015$ ($d$-wave) for $k_z=0$, 
and $\lambda=1.615$ ($s$-wave) and $\lambda=0.377$ ($d$-wave) for $k_z=\pi$.
}
\label{fig:PairJ02D}
\end{figure}
\begin{figure}
\includegraphics[width=\columnwidth]{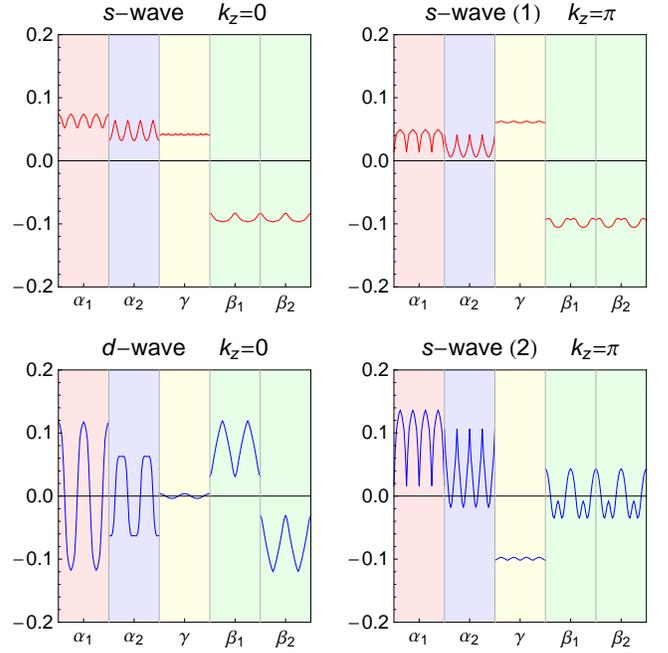}
\caption{(Color online) {\it 2D pairing functions, $\bar{J}>0$:} The leading (upper row) 
and subleading (lower row) pairing functions 
for the hole doped compound ($\langle n \rangle = 5.9$) plotted as before for two different values
of $k_z$, calculated for $\bar{U} = 0.55$ and $\bar{J} = 0.25 \bar{U}$. 
The eigenvalues are 
$\lambda=0.027$ ($s$-wave) and $\lambda=0.015$ ($d$-wave) for $k_z=0$, 
and $\lambda=0.638$ ($s$-wave (1)) and $\lambda=0.381$ ($s$-wave (2)) for $k_z=\pi$.
}
\label{fig:PairJ0252D}
\end{figure}

Again following the notation in Ref.~\onlinecite{ref:Kubo}
we define the singlet pairing vertex in the fluctuation exchange
approximation~\cite{ref:Bickers,ref:Takimoto} as
\begin{eqnarray}
&&\Gamma_{tqps} ({\bf k},{\bf k}',\omega) = \left[\frac{3}{2} U^s
\chi_1^{\rm RPA}({\bf k}-{\bf k}',\omega) U^s + \nonumber \right. \\
&&\left. \frac{1}{2} U^s
 - \frac{1}{2} U^c  \chi_0^{\rm RPA}  ({\bf k}-{\bf k}',\omega)
U^c + \frac{1}{2} U^c \right]_{pstq}
\label{eq:fullGamma}
\end{eqnarray}
where $\chi_1^{\rm RPA}$ denotes the RPA enhanced spin and
$\chi_0^{\rm RPA}$ the RPA enhanced charge susceptibility.
Making use of the Kramers-Kronig relation
we can proceed further by considering only the real part of the
static pairing interaction.
This procedure is justified in the regime 
of weak to intermediate correlations considered herein
and was discussed in detail in a previous work~\cite{ref:GraserNJP}.
If we now confine our considerations to the vicinity of the
Fermi surfaces we can determine the scattering of a Cooper
pair from the state $({\bf k},-{\bf k})$ on the Fermi surface $C_i$ to the state
$({\bf k}',-{\bf k}')$ on the Fermi surface $C_j$ from the projected interaction
vertex
\begin{eqnarray}
{\Gamma}_{ij} ({\bf k},{\bf k}') & = & \mathrm{Re} \sum_{stpq} a_{\nu_i}^{s}({\bf k}) a_{\nu_i}^{t}(-{\bf k}) 
{\Gamma}_{tqps} ({\bf k},{\bf k}',0) \nonumber \\
&& \times a_{\nu_j}^{p,*}({\bf k}') a_{\nu_j}^{q,*}(-{\bf k}') 
\end{eqnarray}
where the momenta ${\bf k}$ and ${\bf k}'$ are restricted to the different
Fermi surface sheets with ${\bf k} \in C_i$  and ${\bf k}' \in C_j$. Defining
a dimensionless pairing strength functional we calculate the
symmetry function $g_\alpha({\bf k})$ of the leading pairing instability from
the following eigenvalue problem
\begin{equation}
- \sum_j \oint_{C_j} \frac{d k_\parallel' d k_z'}{(2\pi)^2} \frac{1}{2\pi v_F(k')} {\Gamma}_{ij} ({\bf k},{\bf k}')
g_\alpha({\bf k}') = \lambda_\alpha g_\alpha({\bf k}) 
\label{EVP}
\end{equation}
where $v_{F}({\bf k}) = |\nabla_{\bf k} E_\nu({\bf k})|$ is the Fermi velocity
on a given Fermi surface. The largest eigenvalue
will lead to the highest transition temperature and its
eigenfunction determines the symmetry of the gap.
For the numerical calculation of the hole doped compound
we parametrize the Fermi surface by a dense
mesh of 1200 ${\bf k}$ values distributed over the 5 different
Fermi surface sheets as shown in Fig.~\ref{fig:FermiS}.

\begin{figure}
\includegraphics[width=\columnwidth]{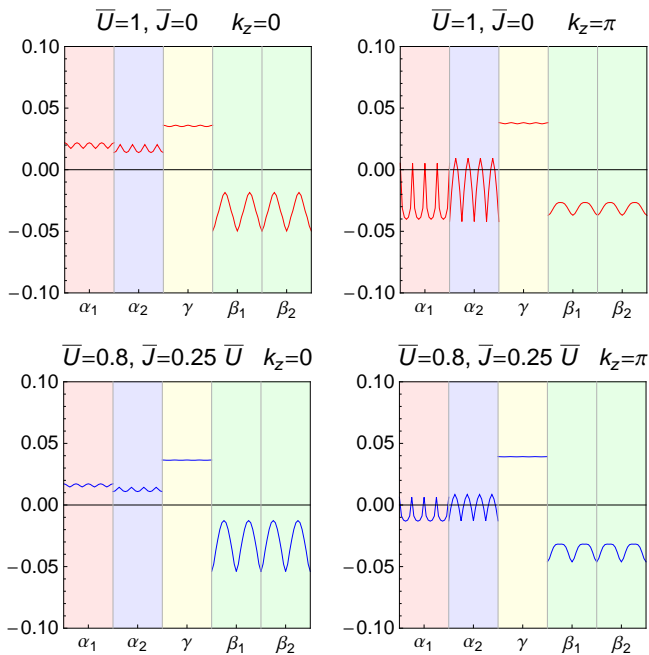}
\caption{(Color online) {\it 3D pairing functions, hole doped:} 
The leading pairing functions 
for the hole doped compound ($\langle n \rangle = 5.9$) plotted as before for two different values
of $k_z$, calculated for $\bar{U} = 1.0$ and $\bar{J} = 0$ (upper row) and  $\bar{U} = 0.8$ 
and $\bar{J} = 0.25 \bar{U}$ 
(lower row). Here the maximum eigenvalues are $\lambda = 0.956$ and $\lambda = 1.077$, respectively.
}
\label{fig:PairJ025}
\end{figure}
\begin{figure}
\includegraphics[width=\columnwidth]{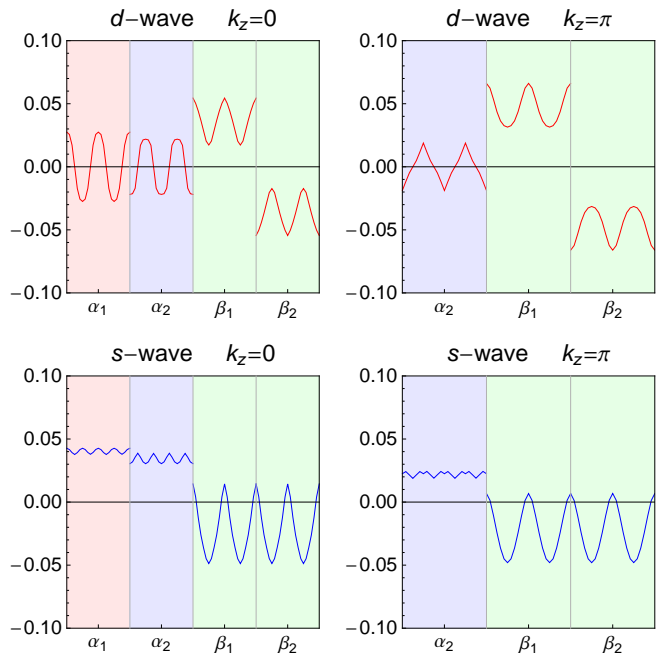}
\caption{(Color online) {\it 3D pairing functions, undoped:} 
The leading and subleading pairing functions 
($d$-wave and $s$-wave)
for the undoped compound ($\langle n \rangle = 6$) plotted as before for two different values
of $k_z$, calculated for $\bar{U} = 0.9$ and $\bar{J} = 0.25 \bar{U}$. Here
the eigenvalues are $\lambda = 1.14$ and $\lambda = 0.617$, respectively.
}
\label{fig:PairJ025x0}
\end{figure}

First we study the pairing function at a fixed
$k_z$ cut of the BZ. In order to solve the eigenvalue problem of
Eq.~\ref{EVP} we first use an effective pairing interaction $\Gamma_{ij}({\bf k},{\bf k}')$ 
calculated from the 2D susceptibility derived in the previous section. 
Due to the different orbital contributions along the Fermi surface 
sheets at $k_z=0$ and $k_z=\pi$ as well as due to the different nesting
conditions, reflected in the respective 2D susceptibilities, 
we find for a given set of interaction parameters quite different 
pairing functions at the center and on top of the BZ. In Fig.~\ref{fig:PairJ02D}
we show the pairing functions for $\bar{U}=0.65$ and $\bar{J}=0$ for a hole doped compound. 
At $k_z=0$ the rather featureless susceptibility (compare Fig.~\ref{fig:2Dsuscept} a),  
without a distinct peak at the nesting vector ${\bf q}=(\pi,0)$, results in a leading 
$s$-wave pairing instability with a sign change between the 
$\alpha$/$\gamma$ and the $\beta$ FS sheets.
At $k_z=\pi$ the strong scattering peak in the susceptibility at ${\bf q}=(\pi,0)$ (Fig.~\ref{fig:2Dsuscept} b) 
drives also an $s$-wave pairing state, but the frustration introduced by the equally 
strong scattering peak at ${\bf q}=(\pi,\pi)$  
enforces an additional sign change between the $\alpha$ and $\gamma$ sheets
and suppresses the gap on the electron pockets.
Comparing the eigenvalues for the two different $k_z$ cuts using the same set of interaction 
parameters we find that we are still far from the instability at $k_z=0$ while we already
have a divergent eigenvalue at $k_z=\pi$. 

If we reduce the intra-orbital pairing interaction $\bar{U}=0.55$ and simultaneously add a finite
Hund's rule coupling $\bar{J}=0.25 \bar{U}$ (Fig.~\ref{fig:PairJ0252D}) we find for both
$k_z$ cuts an $s$-wave symmetry of the leading pairing state 
without the sign change between the $\alpha$ and $\gamma$ Fermi surface sheets at $k_z=\pi$ and a more isotropic
gap size along the electron pockets. This can be 
understood in terms of the enhanced ${\bf q}=(\pi,0)$ peak in the susceptibility without a
simultaneous enhancement of the ${\bf q}=(\pi,\pi)$ peak. 
Here the subleading pairing state is either a $d$-wave ($k_z=0$) or a 
different $s$-wave ($k_z=\pi$).
Obviously a situation where for a fixed set of interaction parameters 
different pairing symmetries at different $k_z$ cuts of the Fermi surface might 
be realized does not reflect an energetically favorable solution: 
thus it is evident that only a full 3D calculation of the pairing state can succeed. 
However, comparing the eigenvalues $\lambda$ for $k_z=0$ and $k_z=\pi$ 
we have confirmed that the primary contributions to the pairing come 
from near $k_z=\pi$.

In a next step we study the pairing function calculated from a full
3D susceptibility using the complete Fermi surface mesh as 
shown in Fig.~\ref{fig:FermiS}. Here the susceptibility is ``averaged''
over $k_z$ and we have already seen that it only weakly depends  
on $q_z$. In the upper row of Fig.~\ref{fig:PairJ025} we show the leading 
eigenfunction for $\bar{U}=1.0$ and $\bar{J}=0$ at $k_z=0$ and $k_z=\pi$ for a 
hole doped compound. Here the 
leading pairing state is an extended $s$-wave state exhibiting  
a higher anisotropy on the electron FS sheets 
$\beta_1$ and $\beta_2$ at $k_z=0$ than at $k_z=\pi$. 
On the hole pockets around $\Gamma$ we find at
$k_z=0$ a small but isotropic gap that develops a strong anisotropy towards $k_z=\pi$,
where it finally exhibits several sign changes around the $\alpha$ FS sheets.
We also observe an overall sign change of the gap on the hole pockets
around $\Gamma$ as a function of $k_z$. 
On the $\gamma$ FS sheet we have a nearly isotropic gap with opposite sign
compared to the $\beta$ sheets in consequence of the pronounced $(\pi,0)$ 
scattering peak in the 3D susceptibility.
For a finite Hund's rule coupling $\bar{J}=0.25 \bar{U}$ (lower row in Fig.~\ref{fig:PairJ025}) 
we find a very similar pairing state with a reduced gap on the $\alpha$ 
Fermi surface sheets at the top of the Brillouin zone.

\begin{figure}[t]
\includegraphics[width=\columnwidth]{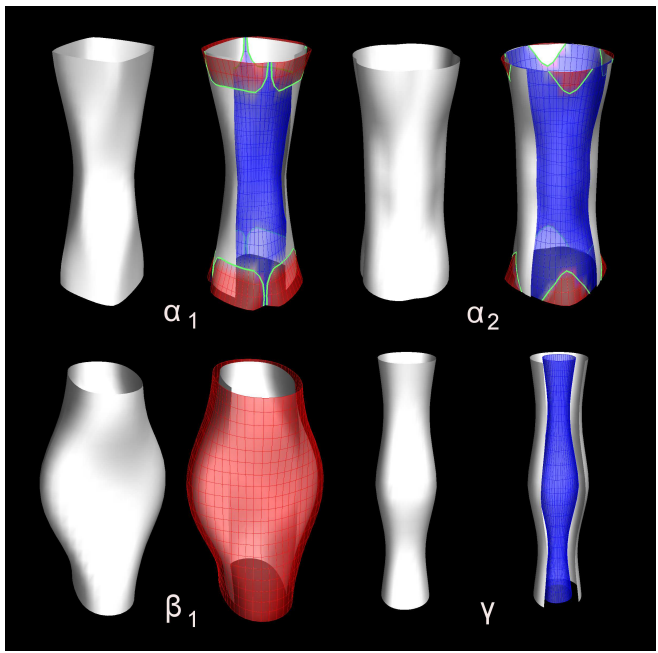}
\caption{(Color online) {\it 3D pairing functions, $\bar{J}=0$:} The leading $A1g$ pairing function 
(extended $s$-wave) on the $\alpha_1$, $\alpha_2$, $\beta_1$, and $\gamma$ Fermi surface sheet
for the hole doped compound ($\langle n \rangle = 5.9$). 
The semi-transparent color mesh visualizes the gap on
each each of the FS sheets that is also shown without the gap
for comparison. The calculations were performed for $\bar{U} = 1.0$ and $\bar{J} = 0$. 
In this figure we have changed the overall phase of the gap 
by -1 from that used in Fig.~\ref{fig:PairJ025} in order to show the nodal structure of the 
gap more clearly.
}
\label{fig:PairJ03D}
\end{figure}

In Fig.~\ref{fig:PairJ03D}, the gap state from the upper row of
Fig.~\ref{fig:PairJ025} is shown along the 
unwrapped Fermi surfaces for the $\alpha_1$, $\alpha_2$, $\beta_1$, and
$\gamma$ sheets. Here it becomes apparent that the sign change of the pairing
state as a function of $k_z$ on the $\alpha_1$ Fermi surface leads to
nearly {\it horizontal} line nodes close to the top of the Brillouin zone.
On the $\alpha_2$ sheet, on the other hand, small V-shaped line nodes
with vertical components are also present.

For the undoped compound (Fig.~\ref{fig:PairJ025x0}) we find a 
$d$-wave solution as the leading eigenfunction for all parameters studied so far.
The anisotropic-$s$ solutions are characterized by greater anisotropy on the 
$\beta$ sheets with nodes in the direction connecting 
the $\alpha$ and $\beta$ FS sheets.
Here we notice that the presence of the pair hopping term $\bar{J}>0$ again
does not change the results qualitatively. This can be understood from the susceptibility 
(Fig.~\ref{fig:3Dsusceptx0}) that does not show a significant enhancement of the
incommensurate scattering peak near ${\bf q} = (\pi,0)$ in the 
presence of a finite pair hopping parameter $\bar{J}$. 
Calculations for an electron doped compound (not shown here)
lead to results that are qualitatively similar to the ones for the undoped compound with a leading
$d$-wave pairing state that is well separated from the subleading sign changing $s$-wave state.

\section{Conclusions}
\label{sec:concl}

In this article we have presented a 5-orbital tight-binding
fit of the DFT band structure of BaFe$_2$As$_2$ derived by unfolding 
the bands of the real BZ into an effective 1Fe/unit cell BZ.
Here we found that the orbital contributions to the Fermi surface sheets
at $k_z=0$ are qualitatively different from the ones at $k_z=\pi$ where 
the hole pockets around $\Gamma$ show a pronounced  
multiorbital composition.
Based on this 5-orbital tight-binding model we compared the 
2D and 3D RPA susceptibilities where the latter is calculated 
by integrating over the full Brillouin zone. We showed that
the susceptibility is dominated by contributions from the top and
the bottom of the BZ and develops well 
pronounced scattering peaks in a certain range of parameters. 
We also found that the $q_z$ dependence of the susceptibility is weak 
but this does not imply that the $k_z$ integration of the susceptibility 
can be neglected since it leads to an averaging of the $k_z$ 
dependent susceptibilities.
In particular the commensurate nature of the 3D magnetic 
response may depend on this averaging.

Finally, we studied the pairing functions in the fluctuation 
exchange approximation and compared again a strictly 2D calculation 
to a complete 3D calculation. Here it becomes obvious that
due to the strong $k_z$ dispersion and the different orbital 
composition of the bands in the center and on top of the 
Brillouin zone the pairing function changes considerably 
along the Fermi surface cylinders and a 2D description will
fail to find the most stable pairing state
over the full Fermi surface. 
Within a 3D approach we showed that for
a moderate hole doping the existence of the additional
hole pocket around $(\pi,\pi)$ favors an 
extended $s$-wave state over the $d$-wave state found for 
the undoped compound. The pairing strength for these states is found to arise predominantly
from processes with momenta near $k_z=\pi$ due to the $d_{xy}$ orbital character of the
hole sheets in this region. The $s$-wave state exhibits 
a strong anisotropy on the electron pockets at 
$k_z=0$ that is reduced at $k_z=\pi$, while on the hole cylinders 
around $\Gamma$ we have a nearly isotropic gap at $k_z=0$
that develops vertical line nodes and changes its sign towards $k_z=\pi$.
For finite $\bar{J}$ the gap maximum on the hole sheets around $\Gamma$ 
is reduced at $k_z=\pi$.
Independent of $\bar{J}$, the 
gap on the hole cylinder around $(\pi,\pi)$ is large and isotropic 
and nearly independent of $k_z$.

Such states should exhibit responses to external probes quite 
different than the 2D states which have been discussed in 
the literature until now. In particular, the nodes near the 
top of the Brillouin zone will contribute strongly to 
low-temperature $c$-axis  transport, and should produce, 
e.g. a strong linear-$T$ term in the penetration depth 
$\lambda_c(T)$ and a weaker behavior in $\lambda_{ab}$, 
as observed recently by Martin {\it et al.}~\cite{ref:Martinetal}. 
 
We emphasize that our primary purpose in this work has been to 
investigate the novel qualitative aspects of the spin fluctuation 
pairing based on the 122-type Fermi surface compared to the more 
familiar 1111-type, in particular those aspects driven by the 
3D dispersion.  As such, we do not claim to have fully explored 
interaction parameter space, nor to have chosen those  parameters 
appropriate to a particular system, as attempted, e.g. in 
Ref.~\onlinecite{ref:Miyake}. Thus while within the limited parameter sets 
we have investigated we have looked at the effects of hole doping, 
it should be understood that these are changes in electronic structure 
which may be controlled by other variables which influence the 
electronic structure, such as the pnictogen/chalcogen position, 
existence of surfaces, or presence of disorder. 
Further work is necessary to see which extensions of the current 
theory are most vital to make quantitative comparisons with experiment.

\begin{acknowledgments}
This work is supported by DOE DE-FG02-05ER46236 (PJH)
and by DOE/BES DE-FG02-02ER45995 (HPC).
SG acknowledges support by the DFG
through SFB 484 and TRR 80 and DJS and TAM acknowledge the
Center for Nanophase Materials Sciences, which is sponsored
at Oak Ridge National Laboratory by the Division of Scientific
User Facilities, U.S. Department of Energy.
We acknowledge NERSC and the University of Florida High-Performance 
Computing Center for providing computational resources that have 
contributed to the research results reported within this paper.
We appreciate stimulating discussions with 
L.~Benfatto, A.~Bernevig, J.~Deisenhofer, A.~Kampf, J.~Kune\v{s}, 
C.~Martin, D.~Singh, R.~Thomale, and F.~Wang.
\end{acknowledgments}

\appendix

\section{Fitting parameters for the 5-orbital tight-binding model}
\label{sec:app}

Subsequently we show the dispersions of the 5-orbital tight-binding fit of the
BaFe$_2$As$_2$ band structure. Here we note that unfolding the small
2Fe/unit cell BZ into an effective large BZ 
corresponding to a unit cell
with only one single Fe and As and a fractional Ba 
is not as straightforward as 
in the case of the 1111 material. Especially the fact that two consecutive FeAs layers cannot
be mapped by a mere translation in $z$ direction leads to
a dependence of part of the interlayer hopping parameters on the
respective sublattice position, and eventually to the necessity
of introducing additional $k_z$ dispersions.
In the following 1 denotes the $d_{xz}$, 2 the $d_{yz}$, 3 the $d_{x^{2}-y^{2}}$, 4 the $d_{xy}$ and 5 the 
$d_{3 z^2-r^2}$ orbital. The hopping parameters are tabulated in 
Table~\ref{tab:intraorbitalhop} and Table~\ref{tab:interorbitalhop}.
In addition we have the 4 onsite energies measured from the Fermi energy as $\epsilon_{1/2}=0.0987$,
$\epsilon_{3}=-0.3595$, $\epsilon_{4}=0.2078$, and $\epsilon_{5}=-0.7516$.
\begin{widetext}
\begin{eqnarray*}
\xi_{11/22}  & = & 2 t_{x/y}^{11} \cos k_x + 2 t_{y/x}^{11} \cos k_y + 4 t_{xy}^{11} \cos k_x \cos k_y
\pm 2 t_{xx}^{11} (\cos 2 k_x -\cos 2 k_y) + 4 t_{xxy/xyy}^{11} \cos 2 k_x \cos k_y  \nonumber \\
& & + 4 t_{xyy/xxy}^{11} \cos 2 k_y \cos k_x + 4 t_{xxyy}^{11} \cos 2 k_x \cos 2 k_y
+ 4 t_{xz}^{11} (\cos k_x +\cos k_y ) \cos k_z \nonumber \\
& & \pm 4 t_{xxz}^{11} (\cos 2 k_x - \cos 2 k_y) \cos k_z \nonumber \\
\xi_{33}  & = & 2 t_{x}^{33} (\cos k_x + \cos k_y) + 4 t_{xy}^{33} \cos k_x \cos k_y
+ 2 t_{xx}^{33} (\cos 2 k_x + \cos 2 k_y)  \nonumber \\
\xi_{44}  & = & 2 t_{x}^{44} (\cos k_x + \cos k_y) + 4 t_{xy}^{44} \cos k_x \cos k_y
+ 2 t_{xx}^{44} (\cos 2 k_x + \cos 2 k_y) + 4 t_{xxy}^{44} (\cos 2 k_x \cos k_y + \cos 2 k_y \cos k_x) \nonumber \\
& & + 4 t_{xxyy}^{44} \cos 2 k_x \cos 2 k_y  + 2 t_{z}^{44} \cos k_z + 4 t_{xz}^{44} (\cos k_x + \cos k_y) \cos k_z + 8 t_{xyz}^{44} \cos k_x \cos k_y \cos k_z \nonumber \\
\xi_{55}  & = & 2 t_{x}^{55} (\cos k_x + \cos k_y) + 2 t_{xx}^{55} (\cos 2 k_x + \cos 2 k_y) 
 + 4 t_{xxy}^{55} (\cos 2 k_x \cos k_y + \cos 2 k_y \cos k_x) \nonumber \\
& & + 4 t_{xxyy}^{55} \cos 2 k_x \cos 2 k_y 
+ 2 t_{z}^{55} \cos k_z + 4 t_{xz}^{55} (\cos k_x + \cos k_y) \cos k_z \\ \\
\xi_{12}  & = &  4 t_{xy}^{12} \sin k_x \sin k_y + 4 t_{xxy}^{12} (\sin 2 k_x \sin k_y + \sin 2 k_y \sin k_x)
+ 4 t_{xxyy}^{12} \sin 2 k_x \sin 2 k_y + 8 t_{xyz}^{12} \sin k_x \sin k_y \cos k_z \nonumber \\
\xi_{13/23}  & = & 2 i t_x^{13} \sin k_{y/x} + 4 i t_{xy}^{13} \sin k_{y/x} \cos k_{x/y}
    - 4i t_{xxy}^{13} (\sin 2 k_{y/x} \cos k_{x/y} - \cos 2 k_{x/y} \sin k_{y/x}) \nonumber \\
\xi_{14/24}  & = & \pm 2 i t_x^{14} \sin k_{x/y} \pm  4 i t_{xy}^{14} \cos k_{y/x} \sin k_{x/y} \pm 4i t_{xxy}^{14}
\sin 2 k_{x/y} \cos k_{y/x} \pm  4 i t_{xz}^{14} \sin k_{x/y} \cos k_z  - 4 t_{xz}^{24} \sin k_{x/y} \sin k_z \nonumber \\
& &  \pm  8 i t_{xyz}^{14} \cos k_{y/x} \sin k_{x/y} \cos k_z
\pm  8 i t_{xxyz}^{14} \sin 2 k_{x/y} \cos k_{y/x} \cos k_z - 8 t_{xxyz}^{24} \sin 2 k_{x/y} \cos k_{y/x} \sin k_z \nonumber \\
\xi_{15/25}  & = & \pm 2 i t_x^{15} \sin k_{y/x} \mp 4i t_{xy}^{15} \sin k_{y/x} \cos k_{x/y} \mp 8 i t_{xyz}^{15}
\sin k_{y/x} \cos k_{x/y} \cos k_z \nonumber \\
\xi_{34}  & = & 4 t_{xxy}^{34} (\sin 2 k_y \sin k_x  - \sin 2 k_x \sin k_y ) \nonumber \\
\xi_{35}  & = & 2 t_x^{35} (\cos k_x - \cos k_y) + 4 t_{xxy}^{35} (\cos 2 k_x \cos k_y - \cos 2 k_y \cos k_x) \nonumber \\
\xi_{45}  & = & 4 t_{xy}^{45} \sin k_x \sin k_y + 4 t_{xxyy}^{45} \sin 2 k_x \sin 2 k_y 
 + 2 i t_z^{45} \sin k_z + 4 i t_{xz}^{45}  (\cos k_x + \cos k_y) \sin k_z \nonumber
\end{eqnarray*}

\begin{table*}
\caption{The intraorbital hopping parameters used for the DFT fit of the 5
orbital model. \label{tab:intraorbitalhop}}
\renewcommand{\arraystretch}{1.4}
\begin{tabular*}{\textwidth}{@{\extracolsep{\fill}}lccccccccccc}
\hline\hline
$t_{i}^{mm}$  & $i=x$ & $i=y$ & $i=xx$  & $i=xy$  & $i=xxy$ & $i=xyy$  & $i=xxyy$  & $i=z$  & $i=xz$ & $i=xxz$ & $i=xyz$ \tabularnewline
\hline
$m=1$  & $-0.0604$ & $-0.3005$  & $0.0253$ & $0.2388$  & $-0.0414$ & $-0.0237$ & $0.0158$ & & $-0.0101$ & $0.0126$ & \tabularnewline
$m=3$  & $0.3378$ &  & $0.0011$ & $-0.0947$ &  &  &  &  &  &  & \tabularnewline
$m=4$  & $0.1965$ &  & $-0.0528$ & $0.1259$ & $-0.032$ &  & $0.0045$ & $0.1001$ & $0.0662$ &  & $0.0421$\tabularnewline
$m=5$  & $-0.0656$ &  & $0.0001$ &  & $0.01$ &  & $0.0047$ & $0.0563$ & $-0.0036$ &  & \tabularnewline
\hline\hline
\end{tabular*}
\end{table*}
\begin{table*}
\caption{The interorbital hopping parameters used for the DFT fit of the 5
orbital model. \label{tab:interorbitalhop}}
\renewcommand{\arraystretch}{1.4}
\begin{tabular*}{0.8\textwidth}{@{\extracolsep{\fill}}lcccccccc}
\hline \hline
$t_{i}^{mn}$  & $i=x$  & $i=xy$  & $i=xxy$  & $i=xxyy$  & $i=z$ & $i=xz$  & $i=xyz$  & $i=xxyz$ \tabularnewline
\hline
$mn=12$  &  & $0.1934$ & $-0.0325$ & $0.0158$ &  &  & $-0.0168$ &  \tabularnewline
$mn=13$  & $-0.4224$ & $0.0589$ & $0.0005$ &  &  &  &  & \tabularnewline
$mn=14$  & $0.1549$ & $-0.007$ & $-0.0055$ &  &  & $0.0524$ & $0.0349$ & $0.0018$ \tabularnewline
$mn=15$  & $-0.0526$ & $-0.0862$ &  &  &  &  & $-0.0203$ & \tabularnewline
$mn=24$  &  &  &  &  &  & $0.0566$ &  & $0.0283$ \tabularnewline
$mn=34$ &  &  & $-0.0108$ &  &  &  &  & \tabularnewline
$mn=35$ & $-0.2845$ &  &  $0.0046$ &  &  &  &  & \tabularnewline
$mn=45$ &  & $-0.0475$ &  & $0.0004$ & $-0.019$ & $-0.0023$ & &  \tabularnewline
\hline \hline
\end{tabular*}
\end{table*}
\end{widetext}

\end{document}